\documentclass[prb, twocolumn, showpacs, showkeys,superscriptaddress]{revtex4}%
\usepackage{amsfonts}
\usepackage{amsmath, amssymb}
\usepackage{graphicx}
\usepackage{psfrag}
\usepackage{subfigure}
\newcommand{\nn}{\nonumber} 
\newcommand{\dv}{\frac{3}{4}}
\newcommand{\ev}{\frac{1}{4}}
\newcommand{\eh}{\frac{1}{2}}

\newcommand{\e}{\epsilon}
\newcommand{\ee}{\mathrm{e}}
\newcommand{\w}{\omega}
\newcommand{\g}{\gamma}
\newcommand{\s}{\sigma}
\newcommand{\al}{\alpha}
\newcommand{\T}{ \left({\vec T}_\alpha\right) }

\newcommand{\LL}{\Lambda}
\newcommand{\si}[1]{_\mathrm{#1}}

\newcommand{\isum}{ \mathrm{sum} }

\newcommand{\pd}{  {\phantom{\dagger}}   }

\begin{document}
\title{Non-equilibrium Scaling Properties of a Double Quantum Dot System:
Comparison between Perturbative Renormalization Group
and Flow Equation Approach}
\author{V. Koerting}
\email[author to whom correspondence should be addressed:\ ]
{koerting@nbi.dk}
\affiliation{Department of Physics, University of Basel, Klingelbergstrasse 82, CH-4056 Basel, Switzerland}
\affiliation{Niels Bohr Institute, Universitetsparken, DK-2100
  Copenhagen \O, Denmark}
\affiliation{The Niels Bohr International Academy,
             The Niels Bohr Institute,
             Blegdamsvej 17,
             DK-2100 Copenhagen,
             Denmark}
\author{P. Fritsch}
\author{S. Kehrein}
\affiliation{Physics Department, Arnold Sommerfeld Center for Theoretical Physics and Center for NanoScience, \\
Ludwig-Maximilians-Universit\"at, Theresienstrasse 37, 80333 Munich,
Germany}


\begin{abstract}

Since the experimental realization of Kondo physics in quantum dots,
its far-from-equilibrium properties have generated considerable theoretical interest.
This is due to the interesting interplay of non-equilibrium physics and correlation effects
in this model, which has by now been analyzed using several new theoretical methods 
that generalize renormalization techniques to non-equilibrium situations.
While very good agreement between these methods has been found for the spin-1/2 Kondo model,
it is desirable to have a better understanding of their applicability for more complicated
impurity models.
In this paper the differences and commons between two such approaches, namely the flow equation
method out of equilibrium and the frequency-dependent poor man's
scaling approach are presented for the non-equilibrium double quantum dot
system. This will turn out to be a particularly suitable testing ground
while being experimentally interesting in its own right.
An outlook is given on the quantum critical behavior of the
double quantum dot system and its accessibility with the two methods.

\end{abstract}
\date{\today}
\maketitle

\section{Introduction}

The recent advances in nanotechnology permit to probe far-from-equilibrium transport
properties of correlated electron systems. The paradigm for such experiments is the
realization of Kondo physics in Coulomb blockade quantum dots \cite{CoulombDots}. This combination of
non-equilibrium with correlation effects is theoretically challenging and has led
to considerable efforts to develop suitable theoretical tools. The key challenge
is that beyond linear response theory one does in general not know how
to systematically construct the steady current-carrying state, except by following
a difficult real time evolution problem.

During the past five years new powerful methods, both numerical and analytical, 
have been developed and used for investigating non-equilibrium quantum
many-body problems. 
As analytical renormalization group methods have played an enormous role in understanding 
correlation physics in equilibrium, a number of these methods are directly based on 
renormalization ideas and carry them over to non-equilibrium situations: the 
frequency-dependent renormalization group \cite{Rosch:03,Rosch:05}, 
the real time renormalization group \cite{SchoellerRev09,Schoeller:09} and
the flow equation method \cite{Kehrein:05,Fritsch:09,Fritsch:10}. While these methods show very good agreement for the
Kondo model, it is clearly desirable for future applications to understand their
relation and respective advantages in more detail. 

In this paper we address this question for the frequency-dependent renormalization group
and the flow equation method in the case of a more complicated quantum impurity model,
namely for a double-dot system in the Kondo regime. In this system two Kondo dots
are coupled via a spin-spin interaction, which is a setup related to
recent experiments~\cite{experiment1,experiment3}. 
In addition, this model is particularly interesting in the present context for two additional 
reasons: i)~It is known to exhibit an interesting quantum phase transition in equilibrium \cite{2imp}
and one expects non-equilibrium properties to be especially important at quantum phase
transitions. 
ii)~Based on the investigation of the non-equilibrium spin-1/2 Kondo model, one knows
that the decoherence generated by the steady state current plays the key role in
understanding the far-from-equilibrium properties \cite{Kaminski:99,Rosch:03}. Now decoherence enters via two
seemingly very different mechanisms in the frequency-dependent renormalization group
and the flow equation method. For the frequency-dependent renormalization group one 
identifies suitable Korringa-like decoherence rates that are then used to explicitly
cut off the renormalization flow \cite{Rosch:03,Paaske:04}. On the other hand, in the flow equation approach decoherence terms appear as two-loop contributions in the scaling
equations \cite{Kehrein:05,Fritsch:09}. For the conventional 
spin-1/2 Kondo system the different renormalization-based approaches show very
good quantitative agreement including line-shapes and Korringa 
rates \cite{Fritsch:09,Fritsch:10,Schoeller:09}.

The difference how decoherence enters highlights the respective advantages 
of these approaches:
the frequency-dependent renormalization group (like real time RG) has a straightforward diagrammatic
representation, while the flow equation method treats both many-particle coherence
effects (like Kondo physics) and decoherence on the same footing in the scaling equations. 
In the following
we will analyze how these differences manifest themselves in the non-equilibrium 
double-dot system. Previous studies of the non-equilibrium double-dot system
based on the frequency-dependent renormalization group and non-equilibrium  
perturbation theory have been published in
Refs.~\onlinecite{Koerting:07,Koerting:08}.

This paper is organized as follows. In section~\ref{sec:Model} we first introduce 
the Hamiltonian of the double
quantum dot system and discuss the various approximations used in its derivation.
Section~\ref{sec:flow} explains the flow equation method and shows its
application to the double dot system. In section~\ref{sec:pRG}
we then introduce the perturbative renormalization group approach and
discuss the scaling equations for our system. 
After comparing the two methods in  leading
logarithmic order in section~\ref{sec:comp}, we discuss generalizations
of the two approaches in section~\ref{sec:beyond} which include
decoherence effects.  At the
end of section~\ref{sec:beyond} we then compare the two approaches again
and show that the results at the decoherence scale are in very
good agreement, although the underlying methodologies are quite different.

\section{Model}
\label{sec:Model}

The simplest model to illustrate the competition between a
spin singlet and a Kondo singlet formation is the
two-impurity Kondo model. It has been studied in detail~Ref.~\onlinecite{2imp} and gained new life with the
progress in nanotechnology and the possibility to study
two single quantum dots interacting with each other~\cite{experiment1,
experiment2, experiment3}.

Here we study the problem of two quantum dots
where two artificial impurities are attached to leads,
in contrast to the historical two-impurity model which
contains two impurities embedded in a metal~\cite{2imp}.

The double quantum dot (DQD) model describes two spin-1/2 degrees of freedom
denoted as $\vec{S}_L$ and $\vec{S}_R$,
which are each Kondo coupled to conduction band electrons with an additional
mutual spin exchange interaction $H_{\rm ex} = K \ \vec{S}_L
\vec{S}_R$, which is assumed to be antiferromagnetic $K>0$.
The Hamiltonian of the system is given by
\begin{align}
  \label{eq:6}
  H = H_{\rm leads} + H_{\rm ex} + H_{\rm Kondo}
\end{align}
The conduction band electrons are described by 
\begin{align}
  H_{\rm leads} =& \sum_{j} \sum_{k, \s} \e\si{k, j}
        : c_{k j \s}^\dag c_{k j \s}^\pd \ ,
\end{align}
where the lead index $j$ is specified later on, 
$\e\si{k,j}$ is the energy of the corresponding conduction band electron and
$c_{k j \s}^\dag$ ($c_{k j \s}^\pd$) are the corresponding creation
(annihilation) operators for a conduction electron with momentum $k$
and spin $\sigma$. The notation $: \ldots :$
denotes normal ordering with respect to the non-interacting ground
state.

The Kondo interaction with the leads is
\begin{align}
  H_{\rm Kondo} =& \sum_{j} \sum_{k'k} J_{k'k}^{Lj}
                 : \vec{S}_L \vec{s}_{(k'j)(kj)} : \nn \\
            &+ \sum_{j} \sum_{k'k}  J_{k'k}^{Rj}
                 : \vec{S}_R \vec{s}_{(k'j)(kj)} : .
\label{eq:Hint_-1}
\end{align}
where $J_{k'k}^{Lj}$ ($J_{k'k}^{Rj}$) is the coupling of the left (right)
quantum dot spin to the spin density of the conduction band electrons in the lead $j$
\begin{align}
: \vec{s}_{(k'j)(kj)} :
&= \sum_{\s'\s} \eh \vec{\tau}_{\s\s'} :c_{k' j \s'}^\dag c_{k j \s}^\pd :
\label{eq:def_s}
\end{align}
and $\vec{\tau}$ are the Pauli matrices of a spin-$1/2$.

In the two-impurity model the spin-spin interaction between the Kondo
spins is mediated by the RKKY interaction~\cite{RKKY},
i.e.~for antiferromagnetic coupling generated by two Kondo spin-flip interactions.
The RKKY interaction depends in sign and strength on the distance
between the two impurities and since it is an effective interaction
in $J^2$ it is both retarded and small (at least every reasonable
theory should do so). Note that the RKKY interaction in this
case scales to the same degree as the Kondo interaction
in scaling theory.

On the contrary singlet-triplet states in quantum dots can arise
from other physical effects, for example 
from exchange couplings and/or orbital degeneracies. Therefore
the effective spin-spin interaction between the Kondo
impurities can be tuned independently from the Kondo interaction
with the leads~\cite{experiment1}.

For both methods we therefore include the spin exchange interaction in
the unperturbed Hamiltonian
\begin{align}
  H_0 =& H_{\rm leads}+H_{\rm ex}
      \label{eq:flow_H0_initial} ,
\end{align}
and we treat the Kondo interaction $H_{\rm Kondo}$ as a small perturbation. 
The eigenstates
of the unperturbed double dot spin system are singlet $\mid 0, 0 \rangle$
with a total spin $S=0$
and triplet states $\mid 1, m \rangle$ with a total
spin of $S=1$ where $m = \{-1, 0, 1\}$.
The perturbative RG focuses on the flow of a generalized coupling
function and the scaling equation is derived by diagrammatic perturbation theory in the
vertex. In order to do the perturbation theory a pseudo-particle
representation for the spin operators is introduced. 
In the flow equation method the scaling equations are derived from
infinitesimal unitary transformations. These involve mainly the
commutation relations of operators and therefore the spin can be
treated as an operator.  

\subsection{Pseudoparticle representation}

In order to calculate diagrams in perturbation theory
and also in perturbative RG, we introduce the 
pseudo particles $d_\gamma^\dag$
which create a singlet or triplet state $\gamma \in \{ s, t_-,t_0, t_+\}$.
The spin exchange interaction Hamiltonian is thus diagonal in the 
pseudo particle operators,
\begin{align}
  H_{\rm ex} &= - \frac{3}{4} K d_s^\dag d_s + \frac{1}{4} K  \
  \sum_\g d^\dag_{t_\g} d_{t_\g} .
\end{align}

The left and right spin (upper and lower sign, respectively)
can be represented by 
bond operators~\cite{bond_op}:
\begin{align}
S_{L/R}^z&=\eh(d_{t_+}^\dag d_{t_+} - d_{t_-}^\dag d_{t_-} \pm d_s^\dag d_{t_0} \pm d_{t_0}^\dag d_s), \\
S_{L/R}^+=\big(S_{L/R}^-\big)^\dag
        &=\eh( d_{t_0}^\dag d_{t_-}^\pd + d_{t_+}^\dag d_{t_0}^\pd 
                                  \pm d_s^\dag d_{t_-}^\pd \mp d_{t_+}^\dag d_s^\pd). 
\end{align}
The constraint
\begin{align}
  Q = d_s^\dag d_s^\pd + d_{t_+}^\dag d_{t_+}^\pd  
    + d_{t_0}^\dag d_{t_0}^\pd  + d_{t_-}^\dag d_{t_-}^\pd = 1, 
  \label{eq:constraint}
\end{align}
is fulfilled by calculating the physical observable from 
the expectation value
\begin{align}
  \langle {\cal O} \rangle_{Q = 1} &= 
      \lim_{\lambda \rightarrow \infty} 
      \frac{\langle Q {\cal O} \rangle_\lambda}{\langle Q \rangle_\lambda},  
\end{align}
where $\langle \ldots\rangle_\lambda$ is the
average over the Hamiltonian $H_\lambda = H + \lambda Q$
where the constraint enters as a chemical potential which
is set to infinity at the end of the calculation~\cite{Abrikosov}.

It is convenient to introduce a matrix representation of
the Kondo spins in the bond operator notation by 
defining a generalized Pauli matrix $\vec{T}_{\alpha}$
leading to 
\begin{align}
  \vec{S}_\alpha = \sum_{\g'\g} \eh d_{\gamma'}^\dag \big( \vec{T}_\alpha
  \big)_{\gamma'\gamma} d_{\gamma}^\pd \; .
\end{align}
In the case of the exchange coupled double quantum dot system this
generalized Pauli matrices are given by
\begin{align}
\vec{T}_L^z&=\begin{pmatrix}0&0&1&0\\0&1&0&0\\1&0&0&0\\0&0&0&-1\end{pmatrix}, 
\qquad
\vec{T}_R^z=\begin{pmatrix}0&0&-1&0\\0&1&0&0\\-1&0&0&0\\0&0&0&-1\end{pmatrix}, 
\\
\vec{T}_L^+&=\big(\vec{T}_L^-)^\dag=\begin{pmatrix}0&0&0&1\\-1&0&1&0\\0&0&0&1\\0&0&0&0\end{pmatrix}, 
\\
\vec{T}_R^+&=\big(\vec{T}_R^-)^\dag=\begin{pmatrix}0&0&0&-1\\1&0&1&0\\0&0&0&1\\0&0&0&0\end{pmatrix}.
\end{align}
Note that the lower right 3x3 matrix, i.e.~the triplet states,
represents the Pauli matrices for a spin-1.

Using this notation the interaction Hamiltonian~\eqref{eq:Hint_-1}
is given in the general form
\begin{align}
  H_{\rm int} &= \sum_{\alpha,j=L,R} \sum_{\g'\g}\sum_{k'\s';k\s}
                \ev  J^{\alpha,j}_{\gamma'\gamma}
                \T_{\g'\g} {\vec \tau}_{\sigma' \sigma} \nonumber 
\\ &\qquad \qquad
\times                d_{\g'}^\dag d_\g^\pd
                :c_{k' j \sigma'}^\dag c_{k j \sigma}^\pd:
\label{eq:Hint_pRG}
\end{align}
During the renormalization we will find that the coupling
between triplet states $J_{tt} d_{t_m}^\dag d_{t_{m'}}^\pd$ 
flows differently than for the spin couplings including
a singlet-to-triplet transition $J_{st} d_{t_m}^\dag d_s^\pd$
and $J_{ts} d_s^\dag d_{t_{m}}^\pd$.

\subsection{Spin notation}

If we do not introduce pseudoparticle states, but
keep the spin operator as a quantity, we find that 
transitions between the eigenstates of the DQD
are given by
\begin{align}
  \left( \vec{S}_L + \vec{S}_R \right)
           |S, m\rangle \rightarrow |S, m \rangle ,\\
\label{eq:spin_P}
  \left( (\vec{S}_L - \vec{S}_R)
         + 2 i (\vec{S}_L \times \vec{S}_R) \right)
           |0, 0\rangle \rightarrow |1, m \rangle ,\\
  \left( (\vec{S}_L - \vec{S}_R)
         - 2 i (\vec{S}_L \times \vec{S}_R) \right)
           |1, m \rangle \rightarrow |0, 0 \rangle .
\label{eq:spin_M}
\end{align}

Therefore we rewrite the Hamiltonian
in Eq.~\eqref{eq:Hint_-1} for the flow equation treatment by
\begin{align}
&  H\si{int} = \sum_{j = L, R} \sum_{k'k}
  J_{k'k}^{\isum, j}
  :\left( \vec{S}_L  + \vec{S}_R \right) \vec{s}_{(k'j)(kj)} :
  \nonumber \\
  &+ \sum_{j = L, R} \sum_{k'k}
  P_{k'k}^{j}
  :\left( (\vec{S}_L  - \vec{S}_R) + 2 i (\vec{S}_L \times \vec{S}_R) \right)
    \vec{s}_{(k'j)(kj)} :
  \nonumber \\
  &+ \sum_{j = L, R} \sum_{k'k}
  M_{k'k}^{j}
  :\left( (\vec{S}_L  - \vec{S}_R) - 2 i (\vec{S}_L \times \vec{S}_R) \right)
    \vec{s}_{(k'j)(kj)} :
  \label{eq:Hint_0} ,
\end{align}
where the couplings are defined by
\begin{eqnarray}
  J^{\isum, j}_{k'k} &=& \eh \left(
                                   J^{L,j}_{k'k}
                                 + J^{R,j}_{k'k}
                                     \right)
 \label{eq:Jsum_initial}, \\
  P^{j}_{k'k} &=& \eh \left( \eh \left(
                                   J^{L,j}_{k'k}
                                 - J^{R,j}_{k'k}
                                     \right)
                                 + Q^j_{k'k}
                              \right)
 \label{eq:P_initial}, \\
  M^{j}_{k'k} &=& \eh \left( \eh \left(
                                   J^{L,j}_{k'k}
                                 - J^{R,j}_{k'k}
                                     \right)
                                 - Q^j_{k'k}
                              \right)
  \label{eq:M_initial},
\end{eqnarray}
The interaction
$Q^j_{k'k}\ :2 i (\vec{S}_L \times \vec{S}_R) \vec{s}_{(k'j)(kj)}:$
is per se not present in the initial
setup, but it turns out that the system of equations
does not close if it is not included.
This leads to the following initial conditions for a general flow
parameter $B$
\begin{align}
  J_{k'k}^{L,j}(B=0) &= J^{L, j}_{k'k} \\
  J_{k'k}^{R,j}(B=0) &= J^{R, j}_{k'k} \\
  Q^j_{k'k}(B=0) &= 0
\end{align}

The following symmetry relations have to be fulfilled
during the flow due to the hermiticity of the Hamiltonian
\begin{eqnarray}
  J^{\isum, j}_{k'k} &=& J^{\isum, j}_{k k'}
 \label{eq:jsum_sym}, \\
  P^{j}_{k'k} &=& M^j_{kk'}
  \label{eq:p_sym}.
\end{eqnarray}
The interaction $P_{k'k}^j$ refers to a scattering process
involving a singlet to triplet transition as indicated
in Eq.~\eqref{eq:spin_P}.
The interaction $M_{k'k}^j$ is the hermitian conjugate of
$P_{k'k}^j = (M_{k'k}^j)^\dag$.

\subsection{Discussion of the leads}

\begin{figure}[htb]
  \centering
  \includegraphics*[width = 0.75 \columnwidth]{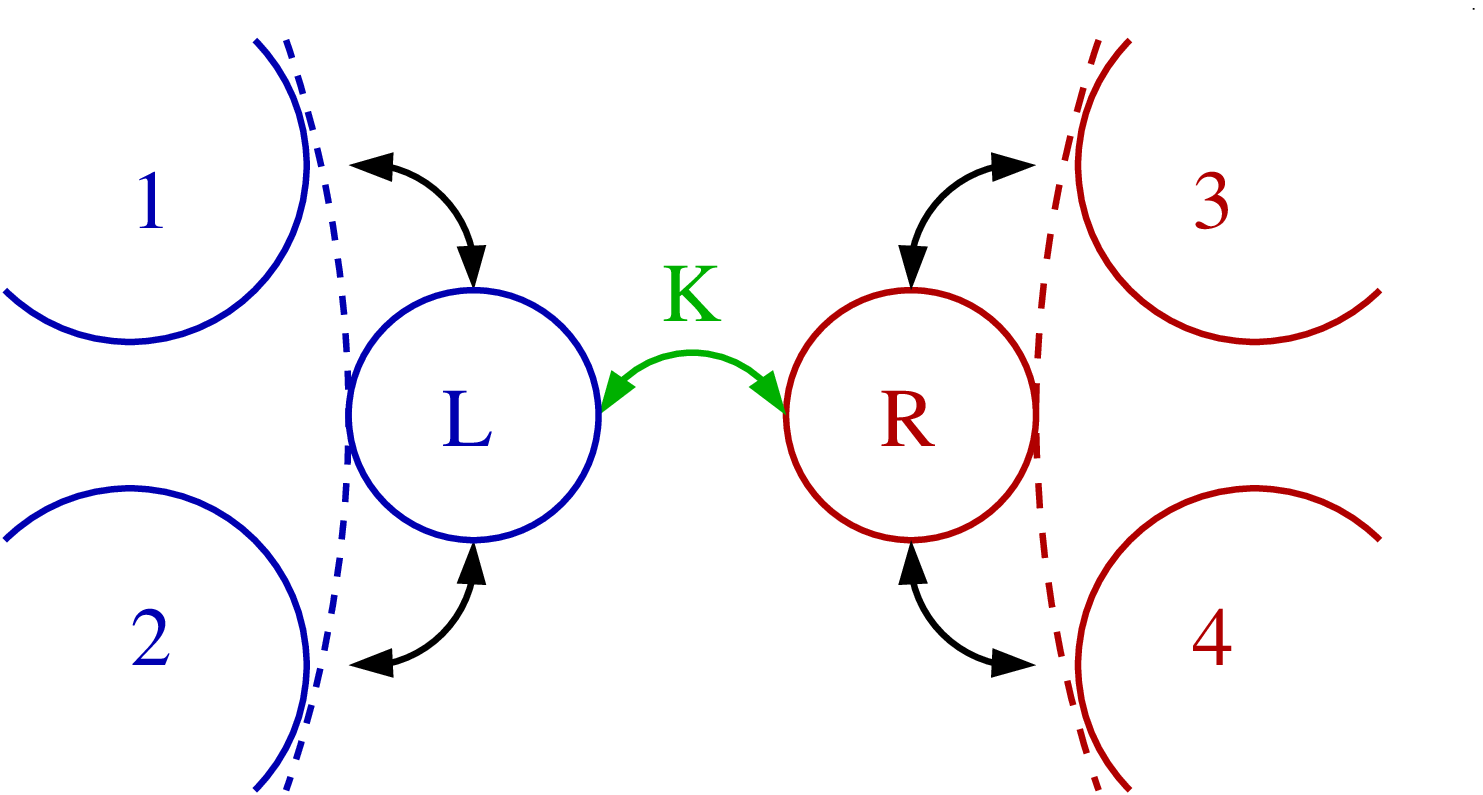}
  \caption{Double Quantum Dot Setup: the residing electrons with a spin
    degree of freedom on the left (L) and right (R)
    quantum dot are coupled mutually by the exchange interaction
    $K$. Two lead $1$ and $2$ ($3$ and $4$) measure the current
    through the left (right) quantum dot. The leads are coupled to the
    quantum dots only by the Kondo spin exchange interaction.
    In the flow equation method we symmetrize
    the leads $1$ and $2$ ($3$ and $4$) to one symmetric left (right)
    lead as denoted by dashed lines.}
  \label{fig:model_device}
\end{figure}

In this paper we will concentrate on the case of
two exchange coupled quantum dots which are not coupled symmetrically to
a set of leads. We want to test the two quantum
dots independently and therefore we assume that there are two leads
attached to each quantum dot such that transport can take
place through each quantum dot independently, see Fig.~\ref{fig:model_device}.

A completely symmetric setup corresponds to two Kondo impurities
embedded in a metal, but in this case
the two impurities couple to the same position in the
lead and for example the RKKY interaction is not defined properly.
On the other hand for completely symmetric coupling
the initial conditions are $J^{\isum,j}_{k'k}(B=0) = J_0$
and $P_{k'k}^j(B=0) = M_{k'k}^j(B=0) = 0$. The singlet-triplet
transitions are not allowed in the beginning and never created
during the flow. Note that $[(S_L+S_R)^2, H] = 0$ and
thus the total spin is conserved. The flow yields
the physics of a non-degenerate singlet or a spin-1 coupled
to leads depending on the initial condition.

In order to make non-equilibrium electron transport possible, two
leads have to be put at different chemical potential, $\mu_{1/2} = \pm eV_L/2$
where $1$ and $2$ denote the two leads attached to the left dot;
similarly $3$ and $4$ for the right two leads, $\mu_{3/4} = \pm
eV_R/2$. While we treat the four leads independently in the pRG approach, we can introduce a simplification due to symmetry arguments in the flow equation calculation.

We focus now for
a short while on the left dot and leads $1$ and $2$.
It has been shown~\cite{Kehrein_book}, that a quantum dot is coupled only
to the even mode of the two leads defined as
\begin{align}
  \label{eq:3}
  c_L = \sqrt{\frac{r_L}{1+r_L}} c_{1} + \sqrt{\frac{1}{1+r_L}} c_{2}
\end{align}
where $r_L = J_{11}/J_{22}$ ($r_R = J_{33}/J_{44}$) is the asymmetry parameter.
For simplicity we only discuss $r_L=r_R=1$ in the numerics. Note that the
extension to $r_\alpha \not=1$ is trivial.
Therefore we have to take into account only $2$ instead of $4$ leads
but with a double step-like occupation function $n_\al(\e)$
\begin{align}
  n_\al(\e) &= \left\{ \begin{matrix}
                     0 & \mathrm{if} & \e &>& eV_\al/2 \\
                     r_\al/(1+r_\al) & \mathrm{if} & |\e| &<& eV_\al/2 \\
                     1 & \mathrm{if} & \e &<& - eV_\al/2
                   \end{matrix}
           \right.
  \label{eq:flow_n_with_voltage}.
\end{align}
where $\al = L, R$ denotes the left or right set of leads.
A non-zero voltage enters the calculation thus via the
normal ordering of the lead electrons, see Ref.~\onlinecite{Kehrein:05}.
In the following we assume that initially no cross-talk
between the left dot and the right leads and vice versa is present,
such that we can "drive" and "probe" the two quantum dots independently by each their leads.
This corresponds to the initial conditions
\begin{align}
  \label{eq:2}
  J^{LR}_{k'k}(B=0) &= J^{RL}_{k'k}(B=0) = 0 \; .
\end{align}
Consequently the initial conditions read
\begin{align}
  \label{eq:inital_parallel}
  J^{LL}_{k'k}(B=0) &= J^{RR}_{k'k}(B=0) = 2 J_0\; , \\
\Rightarrow \qquad
  J^{\isum, L}_{k'k}(B=0) &= J^{\isum, R}_{k'k}(B=0) = J_0\; , \\
  P_{k'k}^{L}(B=0) &= M_{k'k}^L(B=0) = \eh J_0\; , \\
  P_{k'k}^R(B=0) &= M_{k'k}^R(B=0) = - \eh J_0\; ,
\end{align}
where we assumed the symmetry that the left and right coupling
are equal to the fixed but arbitrary value $J_0$.
Note that this model has been studied in detail by one
of the authors in Ref.~\onlinecite{Koerting:07} where
the system showed a current through the left dot even 
when voltage was applied on the right side (transconductance
effect) as discussed in more detail at the end of this work.

\section{Flow equations in lowest order}
\label{sec:flow}

\subsection{The method of flow equations}

In a model with a clear separation of energy scales it is often an
advantage to focus on the low-energy physics and find an effective
representation for the high-energy physics. In the poor man's scaling
approach as will also be discussed in more detail in the next section,
the scattering processes containing energies at the large band edges are
integrated out. Their effective contribution is put into a
renormalized interaction and by further reducing the band cutoff one
arrives at an effective model for a low-energy band.

The separation of energy scales is also important in the flow equation
method. The philosophy here is again to find an effective Hamiltonian
which describes the same physics as the original model but can be
easily solved like a quadratic Hamiltonian. As an example the kinetic
Hamiltonian $\e_{k\s} c^\dag_{k\s} c_{k\s}$ is diagonal in the
conduction electron momenta, whereas the interaction part proportional
to $:\vec{s}_{k'k} = \eh c_{k'\s'}^\dag \vec{\tau}_{\s'\s} c_{k\s}$
connects electrons with different momenta. 
In the matrix representation of the Hamiltonian $H$ we separate
the diagonal contributions, $H_0$, from the off-diagonal contributions $H_{int}$.
The aim of the flow equation method is to generate an effective $H(B)$ starting with
some general flow parameter $B=0$ and modify $H(B)$ accordingly such that
$H(B=\infty)$ is diagonal.

In the flow equation method we achieve this 
by infinitesimal unitary transformations~\cite{Wegner:94}
\begin{align}
  \label{eq:floweq}
  \frac{d}{dB} H(B) &= \big[ \eta(B), H(B) \big].
\end{align}
Note that this expression forces the generator $\eta = - \eta^\dag$ 
to be anti-hermitian which is equivalent to claiming that the
transformation is unitary. The choice
of the generator can be different from problem to problem, but the
canonical generator \cite{Wegner:94}
\begin{align}
  \label{eq:canonical_gen}
  \eta(B) = [ H_0, H_{int}(B) ]
\end{align}
has proven to be a stable choice.
The canonical generator automatically fulfills $\eta^\dag = -
\eta$. Since it is the product of two Hamiltonians it is proportional
to energy$^2$. Consequently the flow parameter $B$ is of the order of
1/energy$^2$.

In contrast to the standard scaling theories this choice of rescaling
eliminates the outermost components in the matrix Hamiltonian or in
other words: scattering processes involving an energy transfer of the
order of the frequency band-cutoff $\LL$ are integrated out in the
course of the flow procedure, $B = 1/\LL^2$. The final Hamiltonian
contains only energy-diagonal processes with a renormalized
energy. It is a clear advantage to keep information on all energy
scales, in particular for the non-equilibrium situation where
scattering processes away from the ground state play an important role.

For a more extended introduction to the flow equation method we 
refer to Ref.~\onlinecite{Kehrein_book}. As a side remark we want to mention
that corrections from taking into account normal ordering with respect to 
the interacting ground state are
of fourth order in the interaction~\cite{Fritsch:09}
and can thus safely be neglected in our calculation to third order.
A first application of the flow equation
method to coupled quantum dots in equilibrium can be found in Ref.~\onlinecite{Garst:04}.

\subsection{Flow equation for the double quantum dot system}

For the double quantum dot system studied in this paper 
the generator $\eta$ is chosen to be the canonical generator
$\eta = [H_0, H_{\rm int}]$ and given explicitly by
\begin{align}
  &\eta = \sum\limits_{j = L,R}  \sum\limits_{k'k}
          \eta_{k' k}^{\isum, j} :(\vec{S}_L + \vec{S}_R) \vec{s}_{(k' j)(k j)}:
  \label{eq:flow_eta_2nd}\\ &
        + \sum\limits_{j = L,R}  \sum\limits_{k'k}
           \eta_{k' k}^{P/M, j}
   :\Big( (\vec{S}_L - \vec{S}_R) \pm 2 i ( \vec{S}_L \times \vec{S}_R )  \Big)
                                 \vec{s}_{(k' j)(k j)}: ,
\notag
\end{align}
where
\begin{eqnarray}
  \eta_{k' k}^{\isum, j} &=&
        \left( \e_{k'} - \e_{k} \right) J_{k'k}^{\isum, j}
  \label{eq:eta_2nd_sum}, \\
 \eta_{k'k}^{P j} &=&
        \left( \e_{k'} - \e_{k} + K \right) P_{k'k}^{j}
 \label{eq:eta_2nd_p} ,\\
  \eta_{k'k}^{M j} &=&
        \left( \e_{k'} - \e_{k} - K \right) M_{k'k}^{j}
  \label{eq:eta_2nd_m} .
\end{eqnarray}
As discussed before we can here observe that the
coupling $P_{k'k}^j$ or $M_{k'k}^j$ corresponds to a transition between
a singlet and triplet state with an energy cost of $\pm K$, respectively.

Due to the construction of the canonical generator ($\propto$ energy$^2$)
the flow parameter $B$ is related to the traditional energy/frequency cutoff $\LL$
by
\begin{align}
  \label{eq:BLambda}
  B \propto \frac{1}{\LL^2}.
\end{align}

Inserting the canonical generator $\eta$ into
the flow equation, Eq.~\eqref{eq:floweq}, we find in lowest, linear order
an exponential behavior of the coupling functions.
Thus we can define an effective coupling $\overline{J_{k'k}^{\isum, j}}(B)$
\begin{align}
  J_{k'k}^{\isum, j}(B) &= \ee^{- B \left( \e_{k'} - \e_k \right)^2} \,
                           \overline{J_{k'k}^{\isum, j}}(B)
  \label{eq:def_diapara_jsum}.
\end{align}
and
\begin{align}
  P_{k'k}^{j}(B) &= \ee^{- B \left( \e_{k'} - \e_k + K \right)^2} \,
                           \overline{P_{k'k}^{j}}(B)
  \label{eq:def_diapara_p}  , \\
  M_{k'k}^{j}(B) = P_{kk'}^j(B) &= \ee^{- B \left( \e_{k'} - \e_k - K \right)^2} \,
                           \overline{P_{kk'}^{j}}(B)
  \label{eq:def_diapara_m}  ,
\end{align}
The effective couplings $\overline{J^{\isum,j}_{k'k}}$ and $\overline{P}^j_{k'k}$
obey a scaling equation with a scaling function $\beta$
which has to be determined from higher than linear order terms.
The exponential dependence though mirrors
the physical picture of the Kondo coupling: It is logarithmically divergent
when energy scattering processes with initial state $k$ and
final state $k'$ are energy-degenerate,
e.g.~$\ee^{- ((\e_{k'} - \e_k)/\LL)^2} = 1$ for $\e_{k'} = \e_k$,
and away from the coherence conditions the coupling functions are suppressed, see also
Fig.~\ref{fig:1loop_exact}
and corresponding discussion.
In contrast to the single-impurity Kondo model (without magnetic
field), a divergent coupling
for $P_{k'k}^j$ representing the singlet-triplet transition can only
be expected when a scattering process in the leads matches the energy
of a transition inside the quantum dot.

The full expression for the flow equation calculation to second order in the
interaction, the so-called one-loop order, is given in the
appendix. In Fig.~\ref{fig:1loop_exact} we 
show full numerical calculations for these
one-loop result in the case of $k'=k$ and compare with  
the solution obtained by the \textit{diagonal} parametrization \cite{Fritsch:09,Fritsch:10}. 
This is a by now well-established approximation that allows some
analytic insight into the flow equations and simplifies the numerical
effort significantly. 

In the diagonal parametrization we assume that the important
energy dependence $\e_k$ (momentum $k$) is given by the exponential decay
$e^{- B (\e_{k'} - \e_k + \al K)^2}$ and we can approximate
\begin{align}
    \label{eq:assumption_dia}
    e^{- B (\e_{k'} - \e_k + \al K)^2} f(\e_{k'}, \e_k) \approx
    e^{- B (\e_{k'} - \e_k + \al K)^2} f(\e_\Sigma) ,
\end{align}
where $\al = \{ 0, \pm 1 \}$ and $\e_\Sigma = (\e_{k'} + \e_k)/2$.

Starting from two energy arguments for the incoming and outgoing
conduction electron one energy is kept fixed but arbitrary and
the other is assumed to fulfill the equation $\e_{k'}- \e_k +\alpha K = 0$.
For example for the coupling to the total spin $(\vec{S}_L +
\vec{S}_R)$ this yields:
\begin{align}
  \label{eq:dia_Jsum_def_more}
  J_{k'k}^{\isum, j}(B) &= \ee^{- B \left( \e_{k'} - \e_k \right)^2} \,
                           J_{(k'+k)/2}^{\isum, j}(B)  \\
\text{where} \quad
  J_k^{\isum, j}(B) &:= J_{k,k}^{\isum, j}(B).
\end{align}
In the diagonal parametrization for $P^j_{k'k}/M^j_{k'k}$
one has to be cautious since the choice 
$\e_{k'} - \e_k = - K$ in $P^j_{k'k}$ is not unique. 
The correct momentum dependence is only recovered if we
choose the diagonal parametrization as
\begin{align}
  \label{eq:dia_P_def_more}
  P_{k'k}^{j}(B) &= \ee^{- B \left( \e_{k'} - \e_k + K \right)^2} \,
                           P_{(k'+k)/2}^{j}(B). \\
\text{where} \quad
  P_k^{j}(B) &:= P_{\epsilon_k-K/2, \epsilon_k+K/2}^j(B)
\end{align}
Note that in the definition of $P_k^j$ the average energy $\e_\Sigma$
is given by
$\e_\Sigma = (\e_{k} - K/2 + \e_k + K/2)/2 = \e_k$.
For the coupling $M^j_{k'k} = P_{kk'}^j$
the assumption in Eq.~\eqref{eq:assumption_dia}, i.e.~$\e_{k'} - K/2 = \e_k + K/2$,
is automatically fulfilled and in diagonal parametrization:
\begin{align}
  \label{eq:dia_M_def_more}
  M_{k'k}^{j}(B) &= \ee^{- B \left( \e_{k'} - \e_k - K \right)^2} \,
                           P_{(k'+k)/2}^{j}(B)  \\
\text{since} & \nonumber \\
  P_k^j(B) &= M_{\e_k+K/2,\e_k-K/2}^j(B) = P_{\e_k-K/2,\e_k+K/2}^j(B).
\end{align}

Using the assumption in Eq.~\eqref{eq:assumption_dia} which leads to
the diagonal parametrization we arrive at the one-loop flow equations
\begin{align}
&\frac{d\, J_{k}^{\isum, j}(B)}{dB} = \nn
\\
& - \sum_{q} (1 - 2 n(qj))
  (\e_{k} - \e_q)
  \ee^{-2 B(\e_{k} - \e_q)^2}
  \big( J_{(k+q)/2}^{\isum, j} \big)^2 \nn
\\ & 
- 4 \sum_{q} (1 - n(qj))
  (\e_{k} - \e_q + K)
  \ee^{-2 B(\e_{k} - \e_q + K)^2}
  \big( P_{(k+q)/2}^{j} \big)^2 \nn
\\ & 
+ 4 \sum_{q} n(qj)
  (\e_{k} - \e_q - K)
  \ee^{-2 B(\e_{k} - \e_q - K)^2}
  \big( P_{(k+q)/2}^{j} \big)^2 \label{eq:j_1loop_1} 
\end{align}
and
\begin{align}
&\frac{d\, P_{k}^{j}(B)}{dB} = \nn \\ 
& - \sum_{q} (1 - n(qj))
  \left( 2 (\e_{k} - \e_q - K/2 ) \right)
  \ee^{- 2 B (\e_{k} - \e_q - K/2 )^2} \nn
\\ & \qquad \qquad
  J_{(k-K/2+q)/2}^{\isum, j}\
  P_{(q+k+K/2)/2}^{j} \nn
\\ & 
+ \sum_{q} n(qj)
  \left( 2 (\e_{k} - \e_q + K/2) \right)
  \ee^{- B (\e_{k} - \e_q + K/2)^2} \nn 
\\ & \qquad \qquad
  P_{(k-K/2+q)/2}^{j}\ 
  J_{(q+k+K/2)/2}^{\isum, j} 
\label{eq:p_1loop_1} 
\end{align}
For details of the calculation we refer to the appendix~\ref{app:1loop}.
Note that $M_{k'k}^j$ is given immediately by the solution for $P_k^j$ in
the diagonal parametrization.

The one-loop order contains the integration over one
internal degree of freedom, the momentum $q$.
Assuming a constant density of states (DOS), $N(0) = 1/(2\LL_0)$, of
a flat band with bandwidth $\LL_0$ around the Fermi energy,
we transform the summation over momenta $q$ to an integral
over the energy $\e_q$:
$\sum_q \to N(0) \int_{-\LL_0}^{\LL_0} d\e_q$.
The DOS is absorbed into the dimensionless
couplings $g^{\isum, j}_{k'k} = N(0) J^{\isum, j}_{k'k}$
and $p^j_{k'k} = N(0) P^j_{k'k}$.

Predicting that the couplings will be only logarithmically dependent
on the energy we
assume that the energy dependence is dominated by the
exponential function and simplify
\begin{align}
  \label{eq:assumption_expstrongdep}
   f(x) \exp(- 2 B (x-c)^2) \approx f(c) \exp(- 2 B (x-c)^2)
\end{align}
Then the integration in Eqs.~\eqref{eq:j_1loop_1} and
\eqref{eq:p_1loop_1} is independent of the energy argument of the
coupling function. 
Straightforwardly we find for example
\begin{align}
&  \int d\e_q (1 - 2 n(qj)) (\e_{k}- \e_q + \alpha K) 
                          \ee^{- 2 B (\e_k - \e_q + \alpha K)^2} \nn \\
=& - \frac{1}{2B} N(0) \Big[ \frac{r_j}{1+r_j} e^{-2B(\e_k - V_j/2 + \alpha
  K)^2} \nn
\\ &\phantom{- \frac{1}{2B} N(0) }
                     + \frac{1}{1+r_j} e^{-2B(\e_k + V_j/2 + \alpha K)^2}                 \Big]
\end{align}
where we used the two-step Fermi function,
Eq.~\eqref{eq:flow_n_with_voltage},  and $\alpha$ 
as placeholder for the corresponding prefactor to $K$.
If we assume that the leads are symmetrically coupled
(asymmetry parameter $r_j = 1$) we can write the
flow equations as
\begin{align}
\frac{d g_{k}^{\isum, j}(B)}{dB} &=
  \frac{1}{2B} \sum_{\nu = \pm 1} \eh\  \ee^{- 2 B (\e_k + \nu V_j/2)^2}
  \left( g_{k}^{\isum, j} \right)^2
\nn \\ & \quad
+ 2 \frac{1}{2B}  \sum_{\nu = \pm 1} \eh\  \ee^{- 2 B (\e_k + K + \nu V_j/2) ^2}
  \left( p_{k+K/2}^{j} \right)^2
\nonumber \\ & \quad 
+ 2 \frac{1}{2B} \sum_{\nu = \pm 1} \eh\  \ee^{- 2 B (\e_k - K + \nu V_j/2)^2}
  \left( p_{k-K/2}^{j} \right)^2
\label{eq:j_1loop_2} 
\end{align}
and
\begin{align}
\frac{d\, p_{k}^{j}(B)}{dB} &=
 \frac{1}{2B} \sum_{\nu = \pm 1} \eh\ \ee^{- 2 B (\e_k - K/2 + \nu V_j/2)^2} \,
  g_{k-K/2}^{\isum, j}\, p_{k}^{j} \nonumber
\\ & \quad
+ \frac{1}{2B} \sum_{\nu = \pm 1} \eh\ \ee^{- 2 B (\e_k + K/2 + \nu V_j/2)^2}\, 
  p_{k}^{j}\, g_{k+K/2}^{\isum, j}
\label{eq:p_1loop_2}
\end{align}
Note that here the term at $\e_k = 0$ is exponentially
small for $B \gg (V_j/2)^2$ due to $\ee^{- B (V_j/2)^2}$ (assuming $K=0$). 
Reducing the band cutoff $\LL$ to $0$ in a system with a large applied
voltage thus leads to a failure of the theory. 
If the voltage is taken beyond the linear response regime 
it is necessary to study the frequency/energy-dependent 
behavior of the coupling functions and the divergence at the two
new Fermi edges $\e_k = \pm V_j/2$.

\subsection{Discussion of results}

Note that the left and right leads do not mix in
Eqs.~\eqref{eq:j_1loop_2}
and \eqref{eq:p_1loop_2} 
and therefore the left
and right coupling can be studied individually.
 
\begin{figure}
 \centering
  \includegraphics*[width = 0.85 \columnwidth]{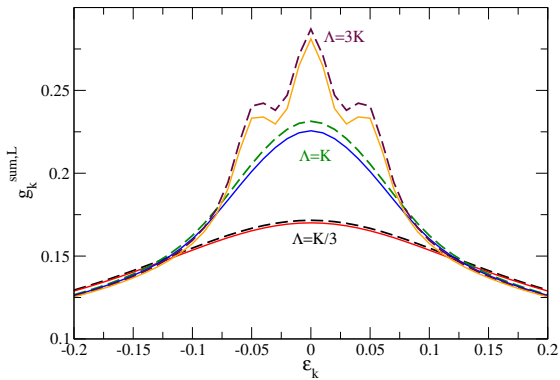}
 \caption{
Flow of the coupling $g^{\isum,L}_{k'k}$ for $\e_{k'} = \e_k$ in the
numerical exact calculation (dashed lines) and for $g^\isum_k$ in the
diagonal parametrization (solid lines)
versus energy $\e_k$ for three different values
$B=1/\LL^2$ and $\LL = K/3, K, 3K$. 
Further parameters of this plot are chosen $K/2 \LL_0 = 0.05$, and
$g(\LL_0) = 0.128$
leading to $T_K = 0.01$.
Note that the flow of $K$ is neglected here.}
\label{fig:1loop_exact}
\end{figure}

In Fig.~\ref{fig:1loop_exact} the flow of the coupling
$g_{k'k}^{\isum, L}$ is plotted for $\e_k = \e_{k'}$ 
versus the energy $\epsilon_k$ for different values of the flow
parameter $B$ while the voltage $V$ is set to zero. 
Since there is no cross-coupling the scaling behavior of
$g_{k'k}^{\isum,R}$ is identical to $g^{\isum,L}_{k'k}$.
The initial value of $g_{k'k}^{\isum, L}$ is energy-independent but 
very soon a frequency dependence is produced, and for large enough $B$
peaks at the Fermi edge $\e_k = 0$ and at non-zero energy $\e_k = \pm
K$ are visible. Values away from $\e_k = 0, \pm K$ are exponentially suppressed. 
As obvious from Fig.~\ref{fig:1loop_exact} the diagonal
parametrization $g_k^{\isum,L}$ reproduces the frequency dependence of
$g_{k'k}^{\isum,L}$ perfectly and also the absolute value of the coupling
is in very good agreement. 
Thus we can conclude from the numerical comparison in
Fig.~\ref{fig:1loop_exact} that the diagonal parametrization is a
good approximation. 

This has proven to be the case for most problems studied previously in the context of
non-equilibrium Kondo models solved with the flow equation
method~\cite{Fritsch:09,Fritsch:10}.
 In the following we will also show that the
diagonal parametrization is equivalent to the non-equilibrium scaling 
method by A.~Rosch et al.~\cite{Rosch:03} and can
thus also confirm the accuracy and the validity of the assumptions in
the other approach.

For $B \ll \mathrm{min}[1/K^2, 1/\e_k^2, 1/V_j^2]$ 
the scaling equation for $g^{\isum,j}_k$ and $p^j_k$ are
identical since the exponential $\ee^{- 2 B (\e_k + \alpha K + \nu V_j/2)^2} \approx 1$
does not cutoff the flow. 
Note though, that the initial value of $p^j_k(B_0) = 1/2\ g_k^{\isum,j}(B_0)$.
Using $B = 1/\LL^2$ where $\LL$ is the frequency cutoff the 
flow equation~\eqref{eq:j_1loop_2} thus simplifies to the well-known Kondo coupling
scaling function for a spin-$1/2$,
\begin{align}
  \frac{dg}{d\ln \LL} = - 2 g^2\; .
\label{eq:kondo1/2}
\end{align}
This equation gives us a one-loop Kondo temperature of 
\begin{align}
  \label{eq:def_TK}
T_K = \LL_0\ \ee^{- 1/2 g(\LL_0)}  \; .
\end{align}
Thus we expect
the flow equation~\eqref{eq:kondo1/2} to diverge and the lowest order to break down
when $B$ reaches the value $1/T_K^2$. Nonzero temperature and
current induced decoherence can remove this divergence as we 
will explain later on. However, first we want to introduce the perturbative
RG approach in the next section.


\section{Generalized perturbative RG}
\label{sec:pRG}

The interaction of a Kondo problem has the general structure
\begin{align}
  H_{\rm int} &= \sum_{\substack{n\sigma,m\sigma' \\ \g,\g' }}
               \ev \T_{\g'\g} {\vec \tau}_{\sigma' \sigma}
  J_{\g, \omega_\g; \g', \omega_{\g'} }
   ^{n \sigma, \omega_c; m \sigma', \omega'_c}
                d_{\g'}^\dag d_\g^\pd
                c_{m \sigma'}^\dag c_{n \sigma}^\pd,
\end{align}
where $\gamma, \gamma'$ refer to the eigenstates of the
eigenstates of the double quantum dot (singlet and triplet in this example),
$m,n$ are the lead indices, 
$\sigma, \sigma'$ are spin up and down states and $\vec \tau_{\sigma'\sigma}$ refers
to the Pauli spin matrix and $\vec T_{\gamma'\gamma}$ is a generalized Pauli matrix which
has to be chosen in accordance with the internal structure of the quantum dot setup.
The momentum dependence of the interaction is neglected and thus we write the
Kondo interaction in terms of the 
momentum integrated
conduction electron operators, $c_{n\sigma} = \sum_{k} c_{nk\sigma}$. 

This Hamiltonian can be furthermore described by a general vertex
\begin{align}
\label{eq:RG_vertex_DQD}
{\cal V}_{\g, \w_\g; \g', \w_{\g'}}^{n \s, \w_c; m \s', \w'_c}
= \ev {\vec \tau}_{\s'\s} \T_{\g'\g}
  N(0) J_{\g, \w_\g; \g', \w_{\g'}}^{n \s, \w_c; m \s', \w'_c} .
\end{align}
Note that the Hamiltonian necessarily has to be hermitian and thus
\begin{align}
{\cal V}_{\g, \w_\g; \g', \w_{\g'}}^{n \s, \w_c; m \s', \w'_c}
= {\cal V}_{\g', \w_{\g'}; \g, \w_\g}^{m \s', \w'_c; n \s, \w_c}
\end{align}
if all couplings are real numbers.

\subsection{General scaling equation}

The idea of scaling has already been introduced. 
Instead of calculating a perturbative series of a physical quantity,
we do a perturbation in the couplings of the interaction. This 
provides us with results beyond standard perturbation theory, which is
known to fail in the Kondo model.
 
One of the first scaling theories was introduced as ''poor man's'' scaling
by P.~W.~Anderson in Ref.~\onlinecite{Anderson:66} and generalized to
non-equilibrium by A.~Rosch et al.~\cite{Rosch:03}. Importantly, the
physics of the problem at hand, i.e.~the expectation value of any physical observable, needs
to be invariant under scaling. Anderson studied the T-matrix 
for the Kondo model and showed that a scattering into a high-energy
state can be absorbed into a lower-energy setup by including the
process to the bandedge $\LL$ and back into an effective
interaction for a smaller bandwidth $\LL - d\LL$.
There are two contributions equivalent to a scattering process to the
upper band edge (electron = Cooper contribution) and lower band edge (hole =
Peierls contribution)~\cite{Rosch:05}. By successively reducing the band cutoff $\LL$ in
infinitesimal steps $d\LL$ we generate an
effective interaction at low energy scales which is of the same form
as the original one but contains a renormalized Kondo coupling $J(\LL)$. 
The change is described by a scaling equation $dJ(\LL)/d\LL$ and
including only one virtual state at the band edge is referred to as one-loop
order. As the poor man's scaling is also perturbative, a truncation has
to be made and only certain renormalization diagrams are taken into account.

In the traditional poor man's scaling~\cite{Anderson:66} the band cutoff is reduced to
zero, $\LL \to 0$. This is a problem in a non-equilibrium situation 
as emphasized before because energies beyond the ground state play an
important role. In Ref.~\onlinecite{Rosch:03} the approach from Anderson was
generalized to renormalize only one of the two band-cutoffs, e.g.~the
outgoing energy $\hbar \w_c'$ for a
vertex while the incoming energy $\hbar w_c$ is fixed (but arbitrary). In
this context the Cooper and Peierls contribution to the scaling
equation have to be calculated in Keldysh notation and we find~\cite{Koerting:08} that
the leading logarithmic contribution originates from an integral of the
form
\begin{align}
  \label{eq:cooper_contr}
 & \frac{\partial}{\partial \ln \LL}  \int\limits_{-\LL}^\LL d\e
     \frac{1}{x - \e}
     \tanh\left( \frac{\e}{2 T}\right) 
  \approx - \frac{\LL}{x - \LL}
\end{align}
i.e.~the real part of the quantum dot Green's function and the lesser
part of the conduction electron Green's function including the sharp
step of the Fermi function at zero temperature. The external energy $x$ can be 
the exchange energy $K$, magnetic field $h$, chemical
potential $\mu$ or combinations of those.

In the perturbative RG method we approximate $\LL/(\LL-x)$ 
by $\Theta(\LL - |x|)$ since
$\LL/(\LL-x) \approx 1$ for $\LL \gg x$ and $\LL/(\LL-x) \approx - \LL/x \approx 0$
for $x \gg \LL$. At this point we like to state that one major difference between
the two scaling method is the choice of the cutoff-function. 
In the flow equation method the cutoff function
$\ee^{- B x^2}$ is valid in general and therefore we expect a better
resolution in the proximity of the logarithmically enhanced peaks.
For further information on the pRG method we refer the interested reader
to Ref.~\onlinecite{Rosch:05} or \onlinecite{Koerting:08}.

In the general notation
\begin{align}
  \label{eq:Hint_general}
  H_{\rm int} &= \sum_{\g\g'} \sum_{n\s;m\s'}
  {\cal V}_{\g, \omega_\g; \g', \omega_{\g'} }
   ^{n \sigma, \omega_c; m \sigma', \omega'_c}
                d_{\g'}^\dag d_\g^\pd
                c_{m \sigma'}^\dag c_{n \sigma}^\pd,  
\end{align}
one can derive a scaling equation of the form
\begin{widetext}
\begin{align}
  \frac{\partial
         {\cal V}_{\g, \w_\g; \g', \w_{\g'}}^{n \s, \w_c; m \s', \w'_c}}
       {\partial \ln \LL}  
 = \eh \sum_{\lambda = \pm 1}  \sum_{\eta, \nu, s}\nn
   &  \Big(
     {\cal V}_{\eta, \w_\eta ; \g', \w_{\g'}}
             ^{\nu s; \lambda \LL + \mu_\nu; m \s', \w'_c}
    \Theta_{|\w_c + \w_\g - \mu_\nu - \e_\eta|}
     {\cal V}_{\g, \w_\g; \eta, \w_\eta}
             ^{n \s, \w_c; \nu s, \lambda \LL + \mu_\nu}
     \nn \\
  & 
  - {\cal V}_{\eta, \w_\eta; \g', \w_{\g'}}
            ^{n \s, \w_c; \nu s, \lambda \LL + \mu_\nu}
    \Theta_{|\w_{\g'} - \w_c + \mu_\nu - \e_\eta|}
    {\cal V}_{\g, \w_\g; \eta, \w_\eta}
            ^{\nu s, \lambda \LL + \mu_\nu; m \s', \w'_c}
   \Big)
  \label{eq:RG_RGeq_general},
\end{align}
\end{widetext}
where we introduced the notation $\Theta_x = \Theta(\LL - |x|)$.
Eq.~\eqref{eq:RG_RGeq_general} is the generalization of the equations given in Ref.~\onlinecite{Rosch:03}.
A derivation can be found in Ref.~\onlinecite{Koerting:08}.
Every ingoing and outcoming leg of the vertex is assigned with a frequency,
$\omega_c, \omega_c'$ and $\omega_\gamma, \omega_{\gamma'}$ for
the conduction electron and quantum dot pseudo fermions, respectively.

\subsection{pRG for the Double Quantum Dot System}

We now concentrate on the case of the double quantum dot system where
the general vertex is of the form Eq.~\eqref{eq:RG_vertex_DQD}.  
Furthermore we assume that there is no external magnetic field applied and
therefore the setup is spatially invariant, i.e.~the three triplet states $t_+, t_0, t_-$
are degenerate. Like in the flow equation approach there are thus
only three vertices: a triplet-triplet transition without energy cost/gain in the DQD,
a singlet-triplet and a triplet-singlet transition involving such a process.

The general vertex has four frequencies assigned to it, where one frequency is fixed due
to energy conservation, which we impose on the vertex. The quantum dot is described by pseudo-particles $d_\gamma$
which have to obey a constraint. In the following we set
the energies "on-shell", e.g.~the frequency of the incoming particle $\w_\gamma$
is given by the eigenenergy of the state $\gamma$, i.e.~$\e_\g$
neglecting a finite lifetime due to hybridization with the leads:
 \begin{align}
  {\cal V}_{\g, \w_\g; \g', \w_{\g'}}^{n \s, \w_c; m \s', \w'_c}
\approx
  {\cal V}_{\g, \e_\g; \g', \e_{\g'}}^{n \s, \w_c; m \s', \w'_c}
       \label{eq:pRG_onshell_approx}
\end{align}
With the energy conservation the vertex thus only depends on one frequency
which is chosen to be the ingoing frequency in the following
 \begin{align}
  {\cal V}_{\g, \e_\g; \g', \e_{\g'}}^{n \s, \w_c; m \s', \w'_c}
= {\cal V}_{\g \g'}^{n  \s; m  \s'}(\w_c)
       \label{eq:pRG_onefrequl_approx}.
\end{align}

Evaluating now the product of Pauli matrices we arrive
at the three scaling equations for the dimensionless Kondo couplings
$g_{\gamma\gamma'}^{nm}(\omega) = N(0) J_{\gamma\gamma'}^{nm}(\omega)$
for zero magnetic field
\begin{align}
 \frac{\partial g_{t  s}^{n  m }( \w ) }
      {\partial \ln \LL}
 = -\eh &\sum_\nu \left(
      2 g_{t  s}^{\nu  m }( \w )
      g_{t  t}^{n  \nu }( \w )
      \Theta_{\w - \mu_\nu }
    \right. \nn \\ & \left.
    + 2 g_{t  s}^{n  \nu }( \w )
      g_{t  t}^{\nu  m }( \w + K )
      \Theta_{\w - \mu_\nu + K }
    \right) ,
\\
\frac{\partial g_{s  t}^{n  m }( \w ) }
      {\partial \ln \LL}
 = -\eh &\sum_\nu \left(
      2 g_{t  t}^{\nu  m }( \w - K)
      g_{s  t}^{n  \nu }( \w )
      \Theta_{\w - \mu_\nu - K}
    \right. \nn \\ & \left.
    + 2 g_{t  t}^{n  \nu }( \w )
      g_{s  t}^{\nu  m }( \w )
      \Theta_{\w - \mu_\nu }
    \right) ,
\\
\frac{\partial g_{t  t}^{n  m }( \w ) }
      {\partial \ln \LL}
 = -\eh &\sum_\nu \left(
      g_{s  t}^{\nu  m }( \w + K )
      g_{t  s}^{n  \nu }( \w )
      \Theta_{\w - \mu_\nu + K}
    \right. \nn \\ & \left.
    + g_{s  t}^{n  \nu }( \w )
      g_{t  s}^{\nu  m }( \w - K )
      \Theta_{\w - \mu_\nu - K}
    \right. \nn \\ & \left.
    + 2 g_{t  t}^{\nu  m }( \w )
      g_{t  t}^{n  \nu }( \w )
      \Theta_{\w - \mu_\nu }
    \right) .
\end{align}
See reference \onlinecite{Koerting:08} for the details of the derivation.

Note that in order to arrive at this scaling equations the
following approximations had to be made.
First, the cutoff is sent to $0$ on the right hand side of
Eq.~\eqref{eq:RG_RGeq_general}.
Otherwise the imposed energy conservation assumed for the left hand
side is not fulfilled on the
right hand side and the RG equation is not self-consistent.
Second, the pseudo-fermions
describing the quantum dots states are assumed to be on-shell, Eq.~\eqref{eq:pRG_onshell_approx}. This step is not necessary
in the flow equation approach since the impurity spin is 
kept as an operator without introducing pseudo particles.
Third, the energy conservation
on the vertex is enforced and thus we end up with only one energy index (which
is chosen to be the ingoing energy).
Fourth, the approximations in Eqs.~\eqref{eq:assumption_dia}
and \eqref{eq:assumption_expstrongdep} are used in the integration
identical to the approximations used in the flow equation method.

Note that the hermiticity of the Hamiltonian which leads to $p_{k'k} = m_{kk'}$
corresponds to
\begin{align}
  \label{eq:5}
g_{st}^{mn}(\w) = g_{ts}^{nm}(\w - K)  
\end{align}
in the perturbative scaling approach.

In order to be able to compare the two results
we introduce a new symmetrized coupling $\tilde g_{st}^{nm}(\omega)$ 
analogous to the diagonal parametrization of $p_k^j$
which is defined as
\begin{align}
  g_{st}^{nm}(\omega) &= 2 \tilde g_{st}^{nm}(\omega - K/2) \; , \\
  g_{ts}^{nm}(\omega) &= 2 \tilde g_{st}^{nm}(\omega + K/2) \; ,
\end{align}
which thus fulfills the same initial condition as $p_k^j$
\begin{align}
  \tilde g_{st}^{nm}(\omega)|_{\LL=\LL_0} &= \eh g_{st}^{nm}(\omega+K/2)|_{\LL=\LL_0}
= \eh N(0) J_0
\end{align}
and is peaked at $\pm K/2$ instead of at $0,K$ and $-K,0$
where $g_{st}^{nm}$ and $g_{ts}^{nm}$ show resonant features.

With this new definition and inserting the approximations
as mentioned above yields the scaling equations

\begin{align}
 \frac{\partial \tilde g_{s  t}^{n  m }( \w ) }
      {\partial \ln \LL}
 &=-  \sum_\nu \left(
      g_{t  t}^{\nu  m }( \w + K/2 )
      \tilde g_{s  t}^{n  \nu }( \w)
      \Theta_{\w - \mu_\nu - K/2}
    \right. \nn \\  &\left.
    + g_{t  t}^{n  \nu }( \w - K/2 )
      \tilde g_{s  t}^{\nu  m }( \w ) 
      \Theta_{\w - \mu_\nu + K/2}
    \right) ,
 \label{eq:RG_RGequations_DQD_st_2} \\
 \frac{\partial g_{t  t}^{n  m }( \w ) }
      {\partial \ln \LL}
 &= -  \sum_\nu \left(
      2 \tilde g_{s  t}^{\nu  m }( \w + K/2)
      \tilde g_{s t}^{n  \nu }( \w + K/2 )
      \Theta_{\w - \mu_\nu + K}
    \right. \nn \\ &\left.
    + 2 \tilde g_{s  t}^{n  \nu }( \w - K/2 )
      \tilde g_{s t}^{\nu  m }( \w - K/2)
      \Theta_{\w - \mu_\nu - K}
    \right. \nn \\ & \left.
    + g_{t  t}^{\nu  m }( \w )
      g_{t  t}^{n  \nu }( \w ) 
      \Theta_{\w - \mu_\nu }
    \right)
 \label{eq:RG_RGequations_DQD_tt_2}
\end{align}
In most applications a further convenient approximation is used:
in the frequency integral over all coupling functions it is assumed
that the main contribution arises from the value at which 
the cutoff functions vanish. Thus the set of equations reduces
to a parametric set instead of a continuous function.
\begin{align}
 \frac{\partial \tilde g_{s  t}^{n  m }( \w ) }
      {\partial \ln \LL}
 &=-  \sum_\nu \left(
      g_{t  t}^{\nu  m }( \mu_\nu )
      \tilde g_{s  t}^{n  \nu }( \mu_\nu + K/2 )
      \Theta_{\w - \mu_\nu - K/2}
    \right. \nn \\ &\left.
    + g_{t  t}^{n  \nu }( \mu_\nu )
      \tilde g_{s  t}^{\nu  m }( \mu_\nu - K/2)
      \Theta_{\w - \mu_\nu + K/2}
    \right) ,
 \label{eq:RG_RGequations_DQD_st_3} \\
 \frac{\partial g_{t  t}^{n  m }( \w ) }
      {\partial \ln \LL}
 &= - \sum_\nu \left(
      2 \tilde g_{s  t}^{\nu  m }( \mu_\nu - K/2)
      \tilde g_{s t}^{n  \nu }( \mu_\nu - K/2 )
      \Theta_{\w - \mu_\nu + K}
    \right. \nn \\ & \left.
    + 2 \tilde g_{s  t}^{n  \nu }( \mu_\nu + K/2 )
      \tilde g_{s t}^{\nu  m }( \mu_\nu + K/2)
      \Theta_{\w - \mu_\nu - K}
    \right. \nn \\ &\left.
    + g_{t  t}^{\nu  m }( \mu_\nu )
      g_{t  t}^{n  \nu }( \mu_\nu )
      \Theta_{\w - \mu_\nu }
    \right)
 \label{eq:RG_RGequations_DQD_tt_3}
\end{align}
We leave out this step while comparing the expressions
\eqref{eq:RG_RGequations_DQD_st_2} and
\eqref{eq:RG_RGequations_DQD_tt_2} 
directly
with Eqs.~\eqref{eq:j_1loop_2} and \eqref{eq:p_1loop_2}
derived with the flow equation method. However, the latter approximation is 
used in the numerical routines to accelerate the calculation.


\section{Comparison of the two methods to one-loop order}
\label{sec:comp}

In the following we will show that the two methods 
use the same approximations and therefore are identical
to leading logarithmic order.

It is straightforward to see that the couplings
in the two different calculations are related by
$g_{tt}(\omega) \sim g^\isum_k$
and $\tilde g_{st}(\omega) \sim p_{k}$, where the energy
$\e_k$ is to be identified with the frequency (energy) $\w$ ($\hbar \w$).

We rewrite Eq.~\eqref{eq:j_1loop_2} 
and Eq.~\eqref{eq:RG_RGequations_DQD_tt_2}
to study in detail the commons and differences
of the pRG and the flow equation method.
\begin{align}
 \frac{\partial g_{t  t}^{n  m }( \w ) }
      {\partial \ln \LL}
 &= -  \sum_\nu \Big(
      g_{t  t}^{n  \nu }( \w ) 
      g_{t  t}^{\nu  m }( \w )
     \Theta_{\w - \mu_\nu }
   \nn \\  &
+   2\ \tilde g_{s t}^{n  \nu }( \w + K/2 )
     \tilde g_{s  t}^{\nu  m }( \w + K/2)
     \Theta_{\w + K - \mu_\nu}
   \nn \\  &
    + 2\ \tilde g_{s  t}^{n  \nu }( \w - K/2 )
        \tilde g_{s t}^{\nu  m }( \w - K/2)
        \Theta_{\w - K - \mu_\nu }
  \Big)
\\
\frac{d g_{k}^{\isum, j}(B)}{d\ln(B^{-1/2})} &=
  - \frac{1}{2} \sum_{\nu} \Big( 
  \left( g_{k}^{\isum, j} \right)^2
   \ee^{- 2 B (\e_k - \mu_\nu)^2}
\nn \\ & \quad
 + 2 \left( p_{k+K/2}^{j} \right)^2
   \ee^{- 2 B (\e_k + K - \mu_\nu) ^2}
\nonumber \\ & \quad 
 + 2 \left( p_{k-K/2}^{j} \right)^2 
   \ee^{- 2 B (\e_k - K - \mu_\nu)^2} \Big)
\label{eq:flow_gsum_again}
\end{align}
The flow parameter $B$ is related to the frequency cutoff
by $B = \LL^{-2}$ and $\mu_v = \pm V_j/2$. Therefore the equations are
actually identical down to the prefactors. The prefactor $1/2$ in
Eq.~\eqref{eq:flow_gsum_again} stems from the even-odd combination of
the leads where $g^{even} = ( 1 + r ) g^{nm}$ and thus $g^{even} = 2
g^{nm}$ for $r = 1$.

Let us repeat the common approximation before we embark on the
differences. Both methods take into account only
the leading logarithmic order by treating only the lowest
order diagrams which contribute to the scaling. In the
perturbative RG the pseudoparticle energies are taken to be onshell
which corresponds to treating the spin operators without a bath. 
In both methods the level of complexity is reduced by imposing
energy conservation on the vertex, i.e.~the energy of the outgoing
conduction electron is given by the energy of the incoming electron
diminished by eventual inelastic processes inside the dot. Since both
methods keep information on the whole bandwidth and the renormalization
reduces only the bandwidth of the state that one scatters into, they
are destined to treat non-equilibrium physics on different 
energy scales, i.e.~when physics of more than the ground state play a role.

The main difference are the two different
cutoff functions $e^{-B x^2} = e^{-(x/\LL)^2}$ 
versus $\Theta_x = \Theta(\LL-|x|)$.
The step function is an approximation in the perturbative
RG whereas the exponential cutoff arises naturally in the 
flow equation method approach. The form of the
cutoff influences at most the lineshape in the proximity of 
logarithmically divergent couplings, see Fig.~\ref{fig:pRG} and
\ref{fig:flow}. 
Since these are heavily influenced
by higher-order renormalization, their shape in leading logarithmic 
order is not reliable anyway and only the flow at the
corresponding energies is described correctly.
 Note though, that for example in the current or other physical
 observables, it is the average over a voltage window which enters, i.e. the integral over the frequency-dependent couplings. Therefore this difference can play a role when comparing result for the non-equilibrium current.

Furthermore there is a summation over the lead indices in the
perturbative RG. In the flow equation calculation we used the
symmetry that only
the even channel of each lead is coupled to the dot and therefore
the lead index drops out. In general the model can also be studied
without doing the even-odd transformation~\cite{Fritsch:09}, 
e.g.~to study models which
are not equivalent to the Anderson impurity model by the Schrieffer-Wolff
transformation.
As explained in the
introductory part, e.g.~the important transport part $g^{12}$
between two leads on the left side is given
by the asymmetry parameter and $g^{L/R,L} = (1+r_L) g^{L/R;22}$:
$g^{L/R;12} = \sqrt{r_L}/(1+r_L) g^{L/R,L}$. We concentrate for the comparison only on
the symmetric case $r_L=1$ and consequently find that 
$ g^\isum_k = 2 g_{tt}(\omega)$
and $p_{k} = 2 \tilde g_{st}(\omega)$.

The strength of the flow equation method is that the diagonal
parametrization is a convenient choice in order to
find analytical expressions, but 
not a necessary limitation of the method. In practice the numerical
cost limits the evaluation of the flow equations to the diagonal parametrization.

Nevertheless both methods are still bound to break down due to the
strong-coupling behavior of the Kondo correlations. In the following
we will explain how decoherence effects are included in the two
different methods, how they provide an additional cutoff and in the
end compare the two methods again.

\section{Beyond one-loop}
\label{sec:beyond}

\subsection{In perturbative RG}

The Kondo problem at zero temperature is known not to be solvable by perturbation theory.
Even though the renormalization schemes can improve
the limit of validity due to an appropriate summation of diagrams,
the theory still breaks down at the energy scale $\LL = T_K$.
However, in equilibrium at finite temperature $T\gtrsim T_K$ or in non-equilibrium
for sufficiently large voltage bias one expects a well-behaved weak coupling expansion.

The seminal work of A.~Rosch \textit{et al.}~\cite{Rosch:03} on the non-equilibrium Kondo model
achieves this by including a physically motivated cutoff given by the
current-induced noise in the system.
Decoherence is unavoidably present due to 
the non-zero steady current flow through the system, which generates
Johnson-Nyquist current noise.
In other words, the quantum dot states gain a finite lifetime due to elastic or
inelastic cotunneling processes with the leads. Also decay
rates due to external baths can play a role. In general a non-zero decoherence rate $\Gamma$
has to be included in the retarded Green's function. This leads to a
new scaling behavior since the leading logarithmic diagrams changes as
\begin{align}
 & \LL \frac{\partial}{\partial \LL}  \int\limits_{-\LL}^\LL d\e
     \frac{x - \e}{(x - \e)^2 + \Gamma^2 }
     \tanh\left( \frac{\e}{2 T}\right) \\
  &\approx - \LL \frac{x - \LL}{(x - \LL)^2 + \Gamma^2 }
     + \LL \frac{x + \LL}{(x + \LL)^2 + \Gamma^2 }.
\end{align}
Thus the cutoff has to be corrected to be
\begin{align}
  \label{eq:4}
  \Theta_x = \Theta(\LL - \sqrt{x^2 + \Gamma^2})
\end{align}
instead of just $\Theta(\LL - |x|)$.
It was shown perturbatively in Ref.~\onlinecite{Paaske:04} that in non-equilibrium
selfenergy and vertex correction become important to the same
degree. Further studies using different renormalization group methods
confirmed the observation that the non-equilibrium decoherence rate
are determined by transport processes and can be different to the
thermodynamically expected expressions.~\cite{Schoeller:09,
  SchoellerRev09, Korb:07, Pletyukhov:10, Holger}

The statement that decoherence terms have to be included
is equivalent to the failure of
the on-shell assumption since the levels in the quantum dot 
are broadened and thus the spin state on the quantum dot gains a finite lifetime.
To find the correct cutoff one has to
calculate the spin susceptibility and find the
correct Lorentzian shape.

\begin{figure}[h]
  \centering
  \includegraphics*[width=0.9 \columnwidth]{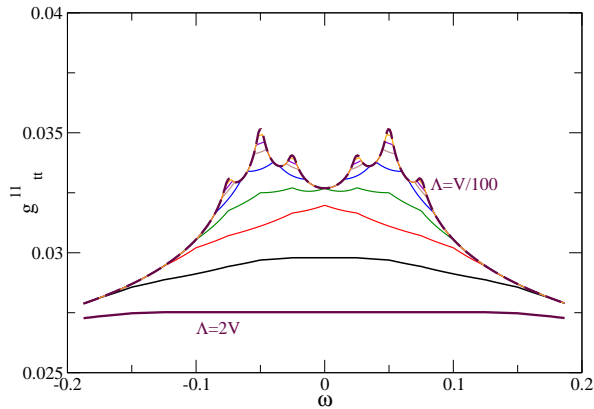}
  \caption{Poor man's scaling including a frequency dependence of the
    coupling for a Double Quantum Dot system; $g_{tt}^{11} = 0.025$, $N(0)
    K = 0.025$, $T\approx0$, $N(0) (eV_L) \equiv V =0.1$, $N(0) (eV_R)
    = 0$; values of the cutoff
    $\LL/V = 2, 1, 1/2, 1/5, 1/10, 1/15, 1/25,1/50,1/100$ where the first
    (solid line)
    and the last value (dashed line) are marked extra in the plot. }
  \label{fig:pRG}
\end{figure}

In Fig.~\ref{fig:pRG} we show the flow of the triplet-triplet coupling
$g^{nm}_{tt}(\omega)$ for $n=m=1$ as a function of the
frequency $\omega$ and for different values of the cutoff $\LL$.

For a large cutoff $\LL$ the coupling is a constant and does not depend
on the frequency of the scattered electrons. As the cutoff is reduced
a frequency dependence evolves and the coupling continues to grow at
the resonant energy scales, i.e.~at $\omega = \pm V/2, \pm K \pm V/2$, while
the flow does not continue at other energy scales due to $\Theta_\w$. Once the energy
scale which is needed for a scattering process is smaller than the
bandwidth, no further processes can be integrated out and it is
expected that the flow should not continue. Including the frequency
dependence and the frequency
cutoff $\Theta_\w$ in the poor man's scaling equations  produces 
this behavior automatically, which is one of the advantages of the pRG method.

Without the cutoff given by $\Gamma$ the couplings at the Fermi energies
$\omega = \pm V/2$ are logarithmically divergent and the poor man's
scaling approach breaks down. Including $\Gamma$ in the cutoff
function $\Theta_\Gamma$ simulates the physical observation that
similar to a non-zero temperature $T \not= 0$ the infinite series of
infinitesimal spin-flip excitations around the Fermi edge is
interrupted by a non-coherent process. The flow of the coupling
stops at $g_{tt} \propto 1/\ln(\Gamma/T_K)$~\cite{Koerting:07}
assuming that $\Gamma$ is the relevant cutoff (which can otherwise
also be proportional to $T$ or combinations of internal energies). 

Note that the divergence
of the coupling at the Fermi edge is not necessarily equivalent to the breakdown
of the calculation. For example in the calculation of the
non-equilibrium current the average of the frequency-dependent
coupling over the whole voltage window needs to be included where the
divergent coupling is just a boundary term~\cite{Holger}.

\subsection{In the flow equation method}

In the flow equation method, the generalization
to higher orders in the coupling function is straightforward as has
already been illustrated in Refs.~\onlinecite{Kehrein:05,Fritsch:09,Fritsch:10}.

In second order in the Kondo coupling $J$ new interaction terms are created.
These can be compared to a 6-leg vertices with
two incoming and two outgoing conduction electrons and
one incoming and one outgoing spin state of the quantum dot
in the pseudo-fermion language. 
Contributions to the flow of the four-leg Kondo interaction vertices
are generated by a product of one of the newly generated vertices with
the initial Kondo vertex. Note that since two conduction electron
lines are thus integrated out, these terms are very similar to the
self energy and vertex corrections discussed in the
previous section.

In contrast to the one-loop contribution,
the two-loop contribution includes an integration
over two intermediate energies, i.e.~an electron-hole pair 
like $n(qv) (1-n(q'v))$. This integration is taken over
a window of energies, e.g.~voltage window or singlet-triplet
transition window. Besides the standard two-loop term
in the $\beta$-function $dg/d\ln \LL = - 2 g^2 + 2 g^3$ for a spin-1
we also expect to find contributions from $g^3 V/\LL$ or $g^3 K/\LL$, i.e.~coupling$^3$
$\times$ phase space. These arise naturally in the calculation
and can be identified as the cutoff rates given by e.g.~the
Johnson-Nyquist current noise. 

\subsubsection{Discussion of the flow to second order}

In the case of a double quantum dot system these newly generated couplings are
\begin{widetext}
\begin{align}
  H_{int}^{(2)}
=& \sum_{jv} \sum_{k'k;q'q} 
   K^{\isum}_{k'k;q'q} i :\big( \vec{S}_L + \vec{S}_R \big)
    \Big( \vec{s}_{(k'j)(kj)} \times \vec{s}_{(q'v)(qv)} \Big): \nn \\
&+ \sum_{jv} \sum_{k'k;q'q} 
   K^{P}_{k'k;q'q} i :\Big( \big( \vec{S}_L - \vec{S}_R \big)
    + 2 i \big( \vec{S}_L \times \vec{S}_R \big) \Big)
    \Big( \vec{s}_{(k'j)(kj)} \times \vec{s}_{(q'v)(qv)} \Big): \nn \\
&+ \sum_{jv} \sum_{k'k;q'q} 
   K^{M}_{k'k;q'q} i :\Big( \big( \vec{S}_L - \vec{S}_R \big)
    - 2 i \big( \vec{S}_L \times \vec{S}_R \big) \Big)
    \Big( \vec{s}_{(k'j)(kj)} \times \vec{s}_{(q'v)(qv)} \Big): \nn \\
& + \sum_{jv} \sum_{k'k;q'q}
  K^{RKKY}_{k'k;q'q} i 2i
  \Big(  :\big( \vec{S}_L \times \vec{s}_{(k'j)(kj)} \big)
    \big( \vec{S}_R \times \vec{s}_{(q'v)(qv)} \big) 
   + \big( \vec{S}_R \times \vec{s}_{(k'j)(kj)} \big)
     \big( \vec{S}_L \times \vec{s}_{(q'v)(qv)} \big): \Big)
\label{eq:Hint2}
\end{align}
\end{widetext}
These interactions are not present in
the initial flow
\begin{align}
   K^{\isum}_{k'k;q'q}(B=0) &= 0 \\
   K^{P}_{k'k;q'q}(B=0) &= 0 \\
   K^{M}_{k'k;q'q}(B=0) &= 0 \\
   K^{RKKY}_{k'k;q'q}(B=0) &= 0
\end{align}
They can be interpreted as an entanglement of the
quantum dot states with the spin states of the
conduction electrons. 
This can be seen by studying of the flow of the coupling
function where we find that the initial free spin state
$(\vec{S}_L + \vec{S}_R)$ evolves into the entangled spin state
$(\vec{S}_L + \vec{S}_R) \times \vec{s}_{(q'v)(qv)}$ for $B \to \infty$.
Correspondingly we observe 
in the flow of the Hamiltonian that the coupling $J_{k'k}$ 
starts to decrease at an energy scale when $K_{k'k;q'q}$ starts to grow.
The flow of $J_{k'k}$ is reversed in the sense that it does not diverge
logarithmically but for $B \to \infty$ we observe $J_{k'k} \to 0$ while the
values of the couplings in $H_{int}^{(2)}$ grow.

Note that  also an RKKY-like interaction is created to this order.
All kinds of potential scattering contributions are neglected here,
similar to the pRG calculation, because
they do not contribute in the wide band limit.
Since we study only antiferromagnetic coupling between the two spins
we can neglect the RKKY interaction in the following. For a
discussion of the flow of this coupling we refer the interested
reader to appendix~\ref{app:RKKY} where it is shown
that the coupling $K$ is similar to a magnetic field in
a spin-$1/2$ Kondo problem renormalized 
by the coupling to the leads~\cite{Fritsch:10}.
Note that the change of $K$ is included in the numerical results shown
here if not stated otherwise.

The rather complicated entanglement term
\begin{align}
&
  K^{RKKY}_{k'k;q'q} i 2i
  \Big(  :\big( \vec{S}_L \times \vec{s}_{(k'j)(kj)} \big)
    \big( \vec{S}_R \times \vec{s}_{(q'v)(qv)} \big) \nn \\
   &+ \big( \vec{S}_R \times \vec{s}_{(k'j)(kj)} \big)
     \big( \vec{S}_L \times \vec{s}_{(q'v)(qv)} \big): \Big) \nn \\
=& 
  (-4)  K^{RKKY}_{k'k;q'q} 
   :\big( \vec{S}_L \vec{S}_R \big)
    \big( \vec{s}_{(k'j)(kj)} \vec{s}_{(q'v)(qv)} \big): \nn \\
  &+ 2  K^{RKKY}_{k'k;q'q}  :\big( \vec{S}_L \vec{s}_{(k'j)(kj)} \big)
    \big( \vec{S}_R \vec{s}_{(q'v)(qv)} \big): \nn \\
   &+ 2  K^{RKKY}_{k'k;q'q} :\big( \vec{S}_R \vec{s}_{(k'j)(kj)} \big)
     \big( \vec{S}_L \vec{s}_{(q'v)(qv)} \big): 
\end{align}
reproduces an RKKY-like interaction.
Note that during the flow
$\vec{S}_L \vec{S}_R \to
(\vec{S}_L \times \vec{s}_{(k'j)(kj)})(\vec{S}_R \times \vec{s}_{(q'v)(qv)})$
and therefore the additional term is effectively
a rescaled exchange interaction. 
This term is not present for example in the spin-1/2 model and special
to the exchange coupled quantum dot. 

We now study systematically how the new terms in the Hamiltonian
decompose into the initial coupling terms. A detailed expression can be
found in the appendix~\ref{app:2loop}. Here we write down only the final
expression for the symmetric setup discussed throughout the paper. 
The flow of the coupling $g^{\isum,j}_k(B)$ in diagonal
parametrization yields
\begin{align}
\frac{d\, g_{k}^{\isum, j}(B)}{dB}
 &= 
\frac{1}{2B} \eh \sum_{v = \pm} \ee^{-2 B (\e_{k} + v V_j/2)^2}
  \big( g_{k}^{\isum, j} \big)^2
\nn \\ &
+ 2 \frac{1}{2B} \eh \sum_{v = \pm} \ee^{-2 B(\e_{k} + v V_j/2 + K)^2}
 \big( p_{k+K/2}^{j} \big)^2
\nn \\ &
+ 2  \frac{1}{2B} \eh \sum_{v = \pm} \ee^{-2 B(\e_{k} + v V_j/2 - K)^2}
 \big( p_{k-K/2}^{j} \big)^2
\nn \\ &
   -  \frac{1}{4B} \sum_v
    g_{k}^{\isum, j}(B)
     \big( g_{max}^{\isum, v}(B) \big)^2
\nn \\ &
   -   \sum_v  f(0, V_v)
    g_{k}^{\isum, j}(B)
     \big( g_{max}^{\isum, v}(B) \big)^2
\nn \\ &
   -  \big( \frac{1}{4 B} \ee^{- 2 B K^2} + \frac{K}{4} \sqrt{\frac{\pi}{2B}} \mathrm{erf}\big(
  \sqrt{2B} K\big) \big)
\nn \\ & \qquad \times
     \sum_v g_{k}^{\isum, j}(B)
   \big( 2   p_{max}^v(B) \big)^2
\nn \\ &
   -  \sum_v f(K, V_v)\ 
    g_{k}^{\isum, j}(B)
   \big(  2 p_{max}^v(B) \big)^2
\label{eq:2loop_gsum}
\; ,
\end{align}
and
\begin{align}
\frac{d\, p_{k}^{j}(B)}{dB} &= \frac{1}{2B}
\eh \sum_{v = \pm } \ee^{- 2 B (\e_{k} + v V_j/2 - K/2)^2}
 g_{k-K/2}^{\isum, j}
  p_{k}^{j}
\nn \\ & \quad
+ \frac{1}{2B}
\eh \sum_{v = \pm} \ee^{- 2 B (\e_{k} + v V_j/2 + K/2)^2}
 p_{k}^{j}
  g_{k+K/2}^{\isum, j}
\nn \\   & \quad
    - \frac{1}{4B}  \sum_v
    p_{k}^{j}(B)
    \big(  g^{\isum, v}_{max}(B) \big)^2
\nn \\   & \quad
    -  \sum_v  f(0, V_v) 
    p_{k}^{j}(B) 
    \big(  g^{\isum, v}_{max}(B) \big)^2
\nn \\   & \quad
     - \big( \frac{1}{4 B} \ee^{- 2 B K^2} + \frac{K}{4} \sqrt{\frac{\pi}{2B}} \mathrm{erf}\big(
  \sqrt{2B} K\big) \big)
\nn \\ & \qquad
\times \sum_v          p_{k}^{j}(B)
          \big( 2 p_{max}^{v}(B) \big)^2
\nn \\   & \quad
     -  \sum_v f(K, V_v)
          p_{k}^{j}(B)
          \big( 2 p_{max}^{v}(B) \big)^2
\; ,
\end{align}
where for $r_L = r_R =1$
\begin{align}
   f(K, V_v) &=  \frac{1}{4 B} \ev \big( \ee^{- 2 B (K-V_v)^2} + \ee^{-
  2 B (K + V_v)^2} - 2 \ee^{- 2 B K^2} \big) \nn \\
&+ \frac{1}{4} \sqrt{\frac{\pi}{2B}} \ev
  \big[ (K-V_v) \mathrm{erf}(\sqrt{2B} (K-V_v))
\nonumber \\ & \qquad
+ (K+V_v) \mathrm{erf}(\sqrt{2B} (K+V_v))
\nonumber \\ & \qquad
- 2 K \mathrm{erf}(\sqrt{2B} K) \big]\; ,
\end{align}
and in particular
\begin{align}
   f(0, V_v) &=  \frac{1}{4 B} \eh \big( \ee^{- 2 B V_v^2}  - 1 \big)
   \nn \\
&+ \frac{1}{8} \sqrt{\frac{\pi}{2B}} 
  V_v \mathrm{erf}(\sqrt{2B} (V_v))\; .
\end{align}

Note that we assumed that the frequency dependence is given 
dominantly by the exponential decay and therefore the couplings in the
integration over the energy window 
can be approximated by their most divergent term,
i.e. $g^{\isum,v}_{(q'+q)/2} \approx g_{max}^{\isum, v} =
g^{\isum,v}_{0}$ if voltage $V_v$ is zero.
For $V_v=0$ there is no contribution from $f(0, 0) = f(K,0) = 0$.
On the contrary for finite voltage and $B \gg V_v^{-2}$: $f(0, V_v) \approx \frac{1}{2B} 
             + \frac{V_v}{2} \sqrt{\frac{\pi}{2B}} \to \frac{V_v}{2} \sqrt{\frac{\pi}{2B}} $
since $\sqrt{\frac{\pi}{2B}} \gg \frac{1}{2B}$ and $f(0, V_v)$ provides
the leading contribution to the flow.

\subsubsection{Two-loop results of the Double Quantum Dot system}

In the equilibrium case, where the applied voltages $V_L = V_R = 0$,
the scaling equation for $g^\isum_k$ is given by
\begin{align}
\frac{d\, g_{k}^{\isum, j}(B)}{dB}
 &\approx 
\frac{1}{2B} \ee^{-2 B\e_{k}^2}
  \big( g_{k}^{\isum, j} \big)^2
\nn \\ &
+ 2 \frac{1}{2B} \ee^{-2 B(\e_{k} + K)^2}
 \big( p_{k+K/2}^{j} \big)^2
\nn \\ &
+ 2  \frac{1}{2B} \ee^{-2 B(\e_{k} - K)^2}
 \big( p_{k-K/2}^{j} \big)^2
\nn \\ &
   -  \frac{1}{4B} \sum_v
    g_{k}^{\isum, j}(B)
     \big( g_{max}^{\isum, v}(B) \big)^2
\nn \\ &
  -  \big( \frac{1}{4 B} \ee^{- 2 B K^2} + \frac{K}{4} \sqrt{\frac{\pi}{2B}} \mathrm{erf}\big(
  \sqrt{2B} K\big) \big)
\nn \\ & \qquad \times
     \sum_v g_{k}^{\isum, j}(B)
   \big( 2   p_{max}^v(B) \big)^2
\label{eq:2loop_gsum_V0}
\; .
\end{align}
Note that the scaling equation is of the form of a two-loop
calculation with an additional contribution proportional to
$(K/\sqrt{2B}) \mathrm{erf}(\sqrt{2B} K)$. For large values of the
argument in the error function we can approximate $\mathrm{erf}(x) \to
\mathrm{sign}(x)$. Therefore we find in the limit $B \to \infty$ that
this term is $\propto 1/\sqrt{B} > 1/B$ and thus this term dominates the asymptotic
flow. As discussed later on, this provides a cutoff in the flow given by 
the singlet-triplet exchange interaction $K$. This cutoff can prevent the
divergence of the elastic Kondo coupling $g^\isum_k$ by favoring a singlet
formation of the two dots instead of a Kondo singlet with the leads if
$K$ is larger than a critical value. This is a signature of the
quantum phase transition in our system and will be
discussed in more detail in another publication~\cite{2012}.

Furthermore we try to illustrate the effect of these decoherence terms
in the two-loop calculation in the limit $V_j \gg K$ and
refer the reader to the explicit discussions in the literature
to non-equilibrium flow equations for further reading~\cite{Kehrein_book}.

Note that we have to distinguish between the voltage applied 
on the left side or right side since there is a summation over the left and
right lead index in third order in the coupling. For now
we will assume $V_R = 0$ and only $V_L \equiv V
\not=0$. In the current this can lead to the interesting effect of a
transconductance~\cite{Koerting:07} as discussed
at the end of this section.


For $2 B V^2 \gg 1 \gg 2 B K^2$
 such that we can set
erf$(\sqrt{2B}V) \approx 1$ and neglect contributions from $\ee^{- 2 B
  K^2}$ we find the scaling equation
\begin{align}
\frac{d\, g_{k}^{\isum, j}(B)}{dB}
 &\approx 
\frac{1}{2B} \sum_{\alpha = \pm 1} \frac{1}{2}\ \ee^{-2 B (\e_{k} + \alpha V/2)^2} 
 \big( g_{k}^{\isum, j} \big)^2
\nn \\ & 
     - \sum_{v = L,R} \Big[ 
        \frac{1}{4} \frac{V_v}{2} \sqrt{\frac{\pi}{2B}}  \Big]
     g_{k}^{\isum, j}(B)
     \big( g_{max}^{\isum, v}(B) \big)^2
\nn     \\ & 
   - \sum_{v=L,R} \Big[ 
         \frac{1}{4} \frac{V_v}{2}  \sqrt{\frac{\pi}{2B}} 
       \Big]
     g_{k}^{\isum, j}(B)
     \big( p_{max}^v(B) \big)^2
\end{align}
In this limit the voltage $V$ dominates the flow for $B \to \infty$ and
thus provides a cutoff scale for the flow.

The actual decoherence rate has to be determined studying the spin
susceptibility of the system. Nevertheless, it has been shown e.g.~for
the spin-1/2 Kondo model~\cite{Fritsch:09} that the rate $\Gamma$ read
off from the correlation function is equivalent to the observed cutoff
in the scaling equation.

\begin{figure}
 \centering
  \includegraphics*[width = 0.9 \columnwidth]{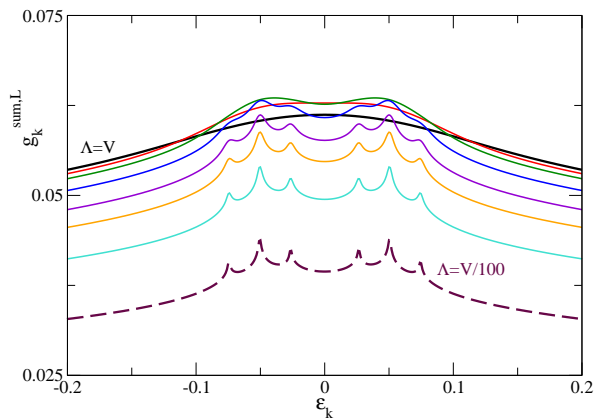}
  \caption{
Flow of $g^{\isum,L}_k$ as a function of energy $\e_k$ for different values of
the flow parameter $B =(1/\LL^2)$
where $\LL$ chosen identical to
Fig.~\ref{fig:pRG}. 
Furthermore $N(0) K = 0.025$ 
$N(0) (eV) = 0.01$, 
$T \approx 0$ and
$g(B_0) = 0.05$. Same parameters as in Fig.~\ref{fig:pRG}
} 
\label{fig:flow}
\end{figure}

In Fig.~\ref{fig:flow} we show (similar to
Fig.~\ref{fig:pRG}) the flow of the coupling $g^{\isum,j}_{k}$ as a
function of $\e_k$ for different values of the flow parameter $B$.
As in the pRG calculation the coupling is initially equal for all values of the
energy and a frequency dependence
evolves slowly with increasing values of $B$. Note the difference
between the two calculations: in the flow equation method the two-loop contribution
leads to a decrease in the running couplings as soon as the energy
scale $\Gamma$ is reached. At this stage the higher order coupling terms
in $H_{int}^{(2)}$ start to grow and the entangled spin
state determines the physics. This is seen as a decrease in the
initial Kondo coupling $g^{\isum,j}_k$, whereas in the poor man's scaling approach
the flow is just stopped at this scale and then stays constant.

It is still an open issue if the coupling exactly at $B = 1/\Gamma^2$
can be used if physical observables are calculated, and if the
cutoff scheme motivated by a self energy cutoff is valid to higher
orders.
For the parameter set chosen in Fig.~\ref{fig:flow} the peak
structure only evolves after the absolute value of the coupling goes
down. Otherwise the overall behavior is very similar to the result
using pRG as shown in Fig.~\ref{fig:pRG}. A direct comparison of
results of the two methods can be
found in the next subsection.

\subsubsection{Transconductance}

A non-zero transconductance is found in the
double quantum dot system~\cite{Koerting:07}.
If the ground state of the double quantum dot system is given by
the degenerate triplet states ($K < 0$), we find a zero-bias resonance which can
be Kondo enhanced. On the other hand if the double quantum dot system 
is in the non-degenerate singlet ground state ($K>0$), 
transport is blocked for voltages below a threshold given by the
exchange interaction $K$. Once the applied voltage is larger than $K$,
transport is also allowed which includes the three triplet states and we
find an inelastic cotunneling step which is logarithmically enhanced by
the Kondo correlations. A
non-equilibrium occupation of the triplet becomes possible due to the large 
applied voltage. For example for a large
voltage applied on the left side the linear response current on
the right side is not blocked but there is a non-zero signal. This
{\em transconductance} $dI_R/dV_L$ was studied in detail by
one of the authors~\cite{Koerting:07, Koerting:08}.

The transfer of decoherence from one quantum dot to another can
also be seen in the flow equation method.
In contrast to one-loop order where only one virtual process is
allowed, there are two intermediate states which contribute to
two-loop order.  Note that there is a summation over the
lead index in Eq.~\eqref{eq:2loop_gsum}. 
Therefore an electron-hole pair created in e.g.~the
right lead due to a finite voltage enters the scaling equation for the
coupling to the left leads.

\begin{figure}[h]
  \centering
  \includegraphics*[width=0.9 \columnwidth]{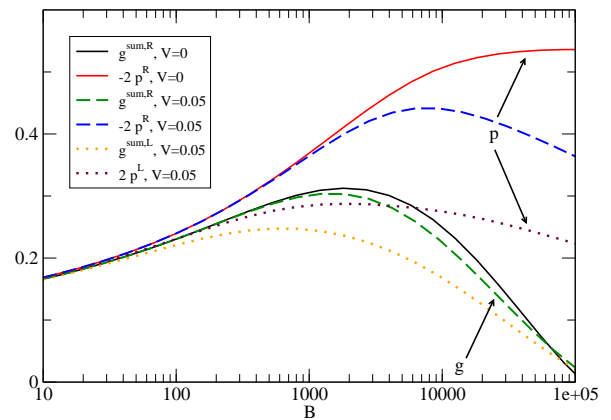}
  \caption{Flow equation solution using the mean value
    of the couplings $g_{mean}^{\isum, R}$ and $- 2 p_{mean}^{\isum,R}$
    to the right leads; solid lines illustrate the flow for voltage
    $V_L=V_R=0$. In this case the singlet-triplet excitation energy
    $K$ provides a cutoff scale in $g^{\isum,j}_{mean}$. For a finite
    voltage applied to the left lead, i.e.~$T_K = 0.005$, $N(0) K = 0.025$,
    $V_R = 0$ and $N(0) (eV_L) \equiv V = 0.05$, we show again $g_{mean}^{\isum, R}$ and $- 2
    p_{mean}^{\isum,R}$ (dashed lines). A decay of the couplings sets
    in which is initiated by the decoherence on the left side of the
    DQD. See discussion of the transconductance effect in the text. 
   For comparison we also show that the decoherence rate for 
   $g_{mean}^{\isum, L}$ and $2 p_{mean}^{\isum,L}$ (dotted lines) is larger. }
  \label{fig:rechts_K_fest}
\end{figure}

The signatures of the transconductance can thus be observed 
as a cutoff of the divergent coupling to the right lead even
when the voltage is only applied to the left lead, 
see Fig.~\ref{fig:rechts_K_fest}.
Note that the decoherence scale though is larger than the
decoherence on the right side. 
To study the transition from a strong-coupling to a weak-coupling
problem, the calculation of either the transconductance or another
physical quantity is necessary. This is not the focus of this paper.
Here we concentrate on a comparison between
the flow equation method and pRG scaling. We have shown that
they are identical to lowest order and by Fig.~\ref{fig:rechts_K_fest} also that 
the flow equation method
provides the same physics as was found from previous studies of the
double quantum dot system using pRG~\cite{Koerting:07}.

\subsection{Comparison}

As a further comparison we show the mean values of
the couplings in the two different calculations.
\begin{figure}[h]
  \centering
  \includegraphics*[width = 0.9\columnwidth]{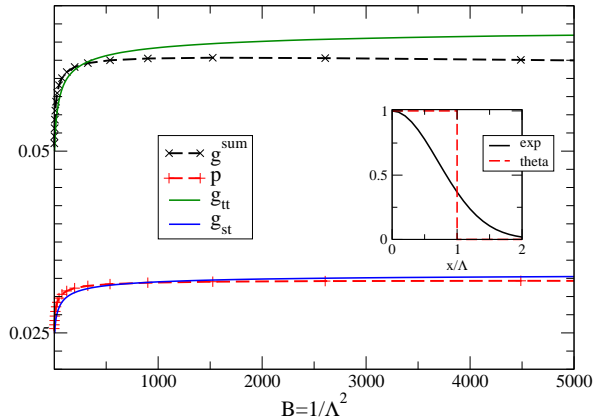}
  \caption{Comparison of the averaged coupling functions in pRG (solid
    lines: $g_{tt} \hat = 2 [g^{11}_{tt}]_{mean}$ and $g_{st} \hat = 2
    [\tilde g^{11}_{st}]_{mean}$) 
    and flow (dashed lines: $g^{sum} \hat = g_{mean}^{sum,L}$ and $p
    \hat= p_{mean}^L$), same parameters as in Fig.~\ref{fig:pRG} and \ref{fig:flow}.
Inset: comparison of the different cutoff functions, $\theta_x$ and
exp$(-(x/\LL)^2)$}
  \label{fig:comp_mean}
\end{figure}
The averaged quantities enter physical quantities
like the current and spin susceptibilities.
They are defined as
\begin{align}
  \label{eq:1}
  g_{mean}^{\isum,j} &= \frac{1}{V_j + 2 K} \int_{-V_j/2 -K}^{V_j/2 +
    K} d\e_k\ g_k^{\isum,j} \\
  p_{mean}^{j} &= \frac{1}{V_j + K} \int_{-V_j/2 -K/2}^{V_j/2 +
    K/2} d\e_k\ p_k^{j}
\end{align}
and analogous for the mean values of 
$g_{tt}^{nm}(\omega)$ and $\tilde g_{st}^{nm}(\omega)$.

In Fig.~\ref{fig:comp_mean} we show a comparison of the 
triplet-triplet and singlet-triplet couplings versus the cutoff/flow
parameter for the same set of parameters as discussed before.
We observe that the couplings in both methods start to grow
logarithmically. The quantitative behavior is slightly different
due to the different cutoff schemes. Note that the plotted
variable is the averaged value over a frequency/energy regime which
thus includes the slopes of the most divergent coherent
couplings, compare Fig.~\ref{fig:pRG} and
\ref{fig:flow}. These are different due to either a cutoff of
$\ee^{- (x/\LL)^2}$ or $\Theta(\LL - x)$ as illustrated in the inset
of Fig.~\ref{fig:comp_mean}. 
Both methods are limited to weak-coupling and thus the
value of $g^\isum_{mean}$ changes only by roughly 25~\% and stays well below
the strong-coupling limit.  

Note that we compare the two-loop flow equations with the
decoherence-cutoff corrected pRG. The philosophies of the two different
methods are obvious in Fig.~\ref{fig:comp_mean}. In the pRG the flow
continues to grow logarithmically and as soon as the reduced band
reaches the cutoff, the flow stops and the coupling stays
constant. This is the value which is then inserted into the expression
of physical observables like the current or the transconductance, etc.
In the flow equation method the decoherence enters differently. As
soon as the flow parameter $B$ reaches $1/\Gamma^2$, the initial couplings
start to decay again. Once the decoherence scale is reached the
Hamiltonian changes its form, i.e.~in the Kondo model the impurity
spin is entangled with the leads and therefore the newly generated
couplings in $H_{int}^{(2)}$ start to grow
and determine the dynamics of the system. For $B \to \infty$ the Kondo
coupling would thus flow to $0$ and the physical relevant value of the
Kondo coupling should be chosen as the maximum before the value starts
to decay. 

The two decoherence scales of the two different methods 
for the example in Fig.~\ref{fig:comp_mean} seem very different since
the pRG flow still continues while the couplings in the flow equation
methods already start to decay.  This observation can be traced back
to two reasons. First, it is not obvious if $B = 1/\LL^2$ is
identically fulfilled or if there is some prefactor involved which
invalidates the direct comparison. 
Second, as pointed out before an
additional cutoff is found in the flow equation method which is 
proportional to the exchange interaction $K$ and not given by the
noise fluctuations. This term is not found from a Korringa-rate
calculation in the pRG method. On the other hand the term contains the
physics of the quantum phase transition in the double quantum dot
system since the Kondo coupling is expected to diverge only if a
Kondo singlet is built up with the conduction electrons in the leads, 
and not if the quantum dot is
locked in a singlet configuration. Detailed study of the behavior of
this transition is the subject of a future publication~\cite{2012}. 

The best comparison which is thus possible between the two methods is 
the frequency-dependent coupling directly at the decoherence scale in
the flow equation method and for $\LL \to 0$ in the pRG approach. 
\begin{figure}[h]
  \centering
  \includegraphics[width = 0.9 \columnwidth]{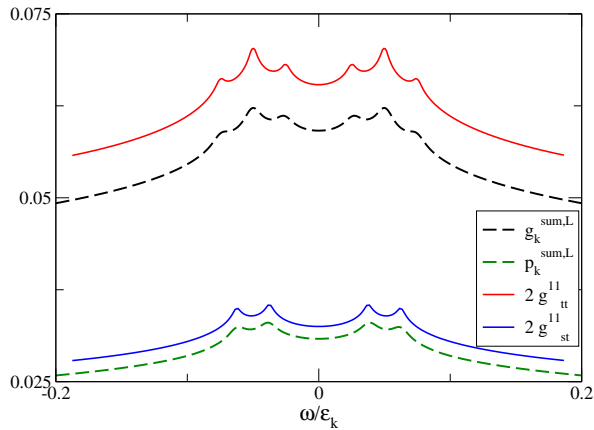}
  \caption{Comparison of the coupling $g_{tt}^{11}(\omega)$ ($\tilde
    g_{ts}^{11}(\omega)$) in the pRG (solid lines) for $\LL \to 0$ 
and $g^{\isum,L}_k$ ($p_k^L$) for the flow equation (dashed lines) at the decoherence
scale $B = 1/\Gamma^2$. 
Parameters are chosen to be the same as in the previous figures. Note
that we compare a 2-loop calculation (flow equation) to an effective
theory which mimics the effects of 2-loop contributions.}
  \label{fig:comp_decoh}
\end{figure}
This is shown in Fig.~\ref{fig:comp_decoh}. The parameters are chosen
such that the voltage is the determining cutoff scale and are
identical to the parameters in all other figures. Note that we only
show the coupling in the lead where the voltage is applied in contrast
to the discussion of the transconductance. We can conclude
that the flow equation method and the pRG approach describe the
same physics of decoherence contributions out of equilibrium.

\section{Conclusion and Outlook}

In conclusion we have shown for the example of the double quantum dot
system that the  flow equation method~\cite{Kehrein:05} and the poor man's scaling
approach to non-equilibrium~\cite{Rosch:03} are equivalent to one-loop order. In both
methods we find that the coupling develops a frequency dependence and
that only at energies/frequencies where coherent processes are
possible a logarithmic divergence of the coupling can be observed.

If the infinite series of coherent processes is broken by decoherence,
e.g.~like in the presence of a not negligible current, the divergence
should be lifted. The two different methods use different approaches
for this: while the pRG approach includes a physically motivated cutoff given
by lifetime broadening and/or vertex corrections, the flow equation
method is continued to higher order, i.e.~two-loop, where a cutoff
arises systematically.

The double quantum dot setup is interesting to study
since it is the simplest model which includes spin-coherent
inelastic scattering processes like the singlet-to-triplet transition at
the exchange energy $K$. The triplet-triplet coupling diverges at the
Fermi energy if the triplets are degenerate, whereas the singlet is
the non-degenerate ground state if $K > 0$. The lifetime of the ground state is
infinite if there is no current applied and thus the double quantum
dot system is also a good case study for the effect of
decoherence due to a finite current. If the current and therefore decoherence
exceed a threshold as discussed in detail in
Ref.~\onlinecite{Koerting:07}, this leads to a non-equilibrium induced current
through the exchange coupled quantum dot even in the linear response
regime. 

A further aspect of the model is the quantum phase transition inherent
in a double quantum system due to the competition between the exchange
coupling induced singlet ground state and the dynamic singlet state
due to the Kondo interaction with the leads. For two impurities
embedded in a metal a quantum phase transition occurs for $K
\approx 2.2 T_K$, which is in a regime where the poor man's scaling
approach breaks down. This quantum phase transition can be studied 
in more detail in the flow equation approach since
i)~the two-loop equation (\ref{eq:2loop_gsum_V0}) 
contains the competition between singlet formation of the two dots
vs. individual Kondo screening
and ii)~the Kondo coupling decreases below the decoherence scale in non-equilibrium. 
In order to do so a physical quantity like the spin susceptibility
will be studied in a different publication~\cite{2012}.
Such a calculation is based on the key feature of the flow
equation approach that decoherence and interaction effects, e.g.
the spin-spin interaction in the double-dot system, are dealt with
on the same footing.

We would like to thank J.~Paaske, P.~W{\"o}lfle and H.~Schoeller
for useful discussions and input at various stages of this project. 
Furthermore we would like to thank
Ch.~Bruder, B.~Braunecker, K.~Flensberg, S.~Andergassen, D.~Schuricht
and L.~Fritz for valuable discussions.

S.~K. acknowledges support through SFB~631 of the
Deutsche Forschungsgemeinschaft (DFG), the Center for Nanoscience (CeNS) Munich, and the German Excellence  Initiative via the Nanosystems Initiative Munich (NIM).

\appendix

\section{One-loop expressions}
\label{app:1loop}

With the expression for $\eta$ in Eq.~\eqref{eq:flow_eta_2nd}
and inserting it into the flow equation  \eqref{eq:floweq},
we find the scaling equation
\begin{widetext}
\begin{align}
\frac{d\, J_{k'k}^{\isum, j}(B)}{dB} &= - (\e_{k'} - \e_k)^2 J_{k'k}^{\isum, j}
\nn \\ & \quad
- \eh \sum_{q} ((1 - n(qj)) - n(qj))
  \left( (\e_{k'} - \e_q) - (\e_q - \e_k)
  \right)  J_{k'q}^{\isum, j} J_{qk}^{\isum, j}
\nn \\ & \quad
- \sum_{q} (1 - n(qj))
  \left( (\e_{k'} - \e_q + K) - (\e_q - \e_k - K)
  \right)  2 P_{k'q}^{j} m_{qk}^{j}
\nn \\ & \quad
+ \sum_{q} n(qj)
  \left( (\e_{k'} - \e_q - K) - (\e_q - \e_k + K)
  \right)  2 m_{k'q}^{j} P_{qk}^{j}
\\
\frac{d\, P_{k'k}^{j}(B)}{dB} &= - (\e_{k'} - \e_k + K)^2 P_{k'k}^{j}
\nn \\ & \quad
- \sum_{q} (1 - n(qj))
  \left( (\e_{k'} - \e_q) - (\e_q - \e_k + K)
  \right)  J_{k'q}^{\isum, j} P_{qk}^{j}
\nn \\ & \quad
+ \sum_{q} n(qj)
  \left( (\e_{k'} - \e_q + K) - (\e_q - \e_k)
  \right)  P_{k'q}^{j} J_{qk}^{\isum, j}
\end{align}
\end{widetext}


With the diagonal parametrization as defined in the main text we can
write
\begin{align}
&\frac{d\, J_{k}^{\isum, j}(B)}{dB} =\nn \\
& - \sum_{q} (1 - 2 n(qj))
  (\e_{k} - \e_q)
  \ee^{-2 B(\e_{k} - \e_q)^2}
  \big( J_{(k+q)/2}^{\isum, j} \big)^2 \nn
\\ & 
- 4 \sum_{q} (1 - n(qj))
  (\e_{k} - \e_q + K)
  \ee^{-2 B(\e_{k} - \e_q + K)^2}
  \big( P_{(k+q)/2}^{j} \big)^2 \nn
\\ & 
+ 4 \sum_{q} n(qj)
  (\e_{k} - \e_q - K)
  \ee^{-2 B(\e_{k} - \e_q - K)^2}
  \big( P_{(k+q)/2}^{j} \big)^2 
\end{align}
\begin{align}
&\frac{d\, P_{k}^{j}(B)}{dB} = \nn \\
& - \sum_{q} (1 - n(qj))
  \left( 2 (\e_{k} - \e_q - K/2 ) \right)
  \ee^{- 2 B (\e_{k} - \e_q - K/2 )^2} \nn
\\ & \qquad \times
  J_{(k-K/2+q)/2}^{\isum, j}
  P_{(q+k+K/2)/2}^{j} \nn
\\ & 
+ \sum_{q} n(qj)
  \left( 2 (\e_{k} - \e_q + K/2) \right)
  \ee^{- B (\e_{k} - \e_q + K/2)^2} \nn 
\\ & \qquad \times
  P_{(k-K/2+q)/2}^{j}
  J_{(q+k+K/2)/2}^{\isum, j}
\end{align}

We assume the exponential dependence on the energy
is stronger than the energy dependence of the coupling function.
Doing the integration assuming that $\ee^{- 2 B (x-c)^2} f(x) \approx
\ee^{- 2 B (x-c)^2} f(c)$
and neglecting contributions from $B_0 = e^{- B \LL_0^2}$ we find
\begin{widetext}
\begin{align}
\frac{d\, g_{k}^{\isum, j}(B)}{dB}
&\approx \quad \frac{1}{2B} \Big(
\frac{r_j}{1+r_j} \ee^{-2 B (\e_{k} - V_j/2)^2} + \frac{1}{1+r_j} \ee^{-2 B
(\e_{k} + V_j/2)^2} \Big)
  \big( g_{k}^{\isum, j} \big)^2
\nn \\ & \quad
+ 2 \frac{1}{2B}
  \Big( \frac{r_j}{1+r_j}\ee^{-2 B(\e_{k} -V_j/2 + K)^2}
  + \frac{r_j}{1+r_j}\ee^{-2 B(\e_{k} -V_j/2 + K)^2}  \Big)
  \big( p_{k+K/2}^{j} \big)^2
\nn \\ & \quad
+ 2  \frac{1}{2B}
  \Big( \frac{r_j}{1+r_j} \ee^{-2 B(\e_{k} - V_j/2 - K)^2}
      + \frac{1}{1+r_j} \ee^{-2 B(\e_{k} + V_j/2 - K)^2} \Big)
  \big( p_{k-K/2}^{j} \big)^2
\\
\frac{d\, p_{k}^{j}(B)}{dB} &= \frac{1}{2B}
  \big( \frac{r_j}{1+r_j} \ee^{- 2 B (\e_{k} - V_j/2 - K/2)^2}
  + \frac{1}{1+r_j} \ee^{- 2 B (\e_{k} + V_j/2 - K/2)^2} \big)
  g_{k-K/2}^{\isum, j}
  p_{k}^{j}
\nn \\ & \quad
+ \frac{1}{2B}
  \big( \frac{r_j}{1+r_j} \ee^{- 2 B (\e_{k} - V_j/2 + K/2)^2}
  + \frac{1}{1+r_j} \ee^{- 2 B (\e_{k} + V_j/2 + K/2)^2} \big)
  p_{k}^{j}
  g_{k+K/2}^{\isum, j}
\end{align}
\end{widetext}
where we have now introduced the dimensionless couplings 
$g^{\isum, j}_{k'k} = N(0) J^{\isum, j}_{k'k}$
and $p^j_{k'k} = N(0) P^j_{k'k}$.
This expression is for $r_j=1$ discussed in detail in the main text.

\section{Flow of the RKKY interaction}
\label{app:RKKY}

In second order in the Kondo coupling we also generate an RKKY-like
spin exchange interaction, $I_{RKKY} \vec{S}_L \vec{S}_R$. 
Together with the flow of $I_{RKKY}$
\begin{align}
\frac{d}{dB}I_{RKKY} 
=& \sum_{k'k} \Big\{ 
[n(k'j) - n(kj)] \eta_{k'k}^{\isum, j} J_{kk'}^{\isum, j}
\nn \\ & \quad
- [n(k'j) - n(kj)] \nn \\ & \qquad
\times \big( 2\ \eta_{k'k}^{Pj} M_{kk'}^j  + 2\ \eta_{k'k}^{Mj} P_{kk'}^j \big)
\nn \\ & \quad
+ [n(kj) (1 - n(k'j)) + n(k'j) (1- n(kj))]
\nn \\ & \qquad
\times \big( 4\ \eta_{k'k}^{Pj} M_{kk'}^j - 4\ \eta_{k'k}^{Mj} P_{kk'}^j \big)
\Big\}
\label{eq:flow_RKKY}
\end{align}
there is also a constant proportional to $3/4 \hbar^2$ generated
in the flow to second order.
\begin{align}
  \frac{d}{dB}E_{const} &= \dv \hbar^2 \sum_{k'k} \Big\{ 
[n(k'j) - n(kj)] \eta_{k'k}^{\isum, j} J_{kk'}^{\isum, j}
 \nn \\ & \quad
+ [n(k'j) - n(kj)]  \big( 2\ \eta_{k'k}^{Pj} M_{kk'}^j + 2\
\eta_{k'k}^{Mj} P_{kk'}^j
 \big)
\Big\}
\end{align}
In the following we study only the case for zero voltages, $V_L = V_R = 0$.

If we now insert the diagonal parametrization for the couplings and do
the energy integration assuming like before that the energy dependence
of the couplings can be neglected. The integration over momentum yields
\begin{align*}
&  \int d\e_{k'} \int d\e_k [n(kj) (1 - n(k'j))] \\ & \qquad \times
     (\e_{k'} - \e_k + \alpha K) \ee^{- 2 B (\e_{k'} - \e_k + \alpha K)^2}\\
&= \frac{\sqrt{2\pi}}{16 B^{3/2}}
   \Big( 1 - {\rm erf}(\sqrt{2B}\alpha K) \Big) \\
&  \int d\e_{k'} \int d\e_k [n(k'j) (1 - n(kj))] \\ & \qquad \times
     (\e_{k'} - \e_k + \alpha K) \ee^{- 2 B (\e_{k'} - \e_k + \alpha K)^2} \\
&= - \frac{\sqrt{2\pi}}{16 B^{3/2}}
   \Big( 1 + {\rm erf}(\sqrt{2B}\alpha K) \Big)
\end{align*}
Utilizing these results we thus find
\begin{align}
  \frac{d}{dB}I_{RKKY} =&   \frac{\sqrt{2\pi}}{8 B^{3/2}}
\Big\{ 
\big[ 1 -2 {\rm erf}(\sqrt{2B}K) \big] ( 2 p_{max}^j )^2 
\\ & \qquad \quad - (g_{max}^{\isum, j})^2 
\Big\}
\end{align}
\begin{align}
  \frac{d}{dB}E_{const} &= \dv \hbar^2 
  \frac{\sqrt{2\pi}}{8 B^{3/2}}
\Big\{ 
-  (g_{max}^{\isum, j})^2 
- ( 2 p_{max}^j )^2 
\Big\}
\end{align}
where $g_{max}^{\isum,j}$ and $p_{max}^j$ are the couplings at the
most divergent energy argument. 

Note that for $BK^2 \gg 1$ where $\mathrm{erf}(\sqrt{2B}K) \approx 1$, the following relation holds
\begin{align}
  \frac{3}{4} \hbar^2 \frac{d I_{RKKY}}{dB} - \frac{d E_{const}}{dB}
  &= 0
\end{align}
i.e.~the combination of the two newly generated couplings stops 
to flow as soon as $B$ has reached the energy
scale $1/K^2$. Corrections to each of the couplings $I_{RKKY}$ and
$E_{const}$ are given by a term proportional to
$B^{- 3/2}$ which is negligible small compared to the logarithmic
divergence for small values of $B$.

Furthermore in the opposite limit $BK^2 \ll 1$
\begin{align}
&\frac{d I_{RKKY}}{dB} 
\nn \\
 &\approx -  \frac{\sqrt{2\pi}}{8 B^{3/2}}
\Big\{ 
\big[ - 1 + 2 \frac{2}{\sqrt{\pi}} \sqrt{2B} K  \big] ( 2 p_{max}^j)^2 
+ (g_{max}^{\isum,j})^2
\Big\} \nn \\
 &\approx -  
 \frac{K}{B}  ( 2 p_{max}^j )^2 
\end{align}
where we neglect the small term $\propto 1/B^{3/2}$.

Note that the frequency dependence of the coupling functions is not
developed in the limit $BK^2 \gg 1$ and thus we find that the relation
$g^{\isum,j}_{max} = 2 p_{max}^j$ is fulfilled during the flow. Thus
for example the coupling $g^{\isum,j}_{max}$ fulfills the flow equation
\begin{align}
  \frac{dg_{max}^{\isum,j}}{dB} &= \frac{1}{B} (g_{max}^{\isum,j})^2
  \\ 
\Rightarrow \qquad
g_{max}^{\isum,j} &= \frac{1}{\ln(B T_K^2)}
\end{align}
where $T_K$ is the Kondo temperature as defined in the main text.
In the following we assume that the initial value $K$ on the right
hand side of the flow equation is also subject of the flow and thus
replace: $K \to I_{RKKY}$. Consequently we have to solve the
differential equation
\begin{align}
d \ln I_{RKKY}
&\approx -  2
\Big( \frac{1}{\ln(BT_K^2)} \Big)^2\ d\ln B 
\end{align}
where the factor of $2$ originates from the summation over the two lead
indices $j$ which is special to the chosen model. The latter equation
can be solved immediately and yields
\begin{align}
I_{RKKY}(B) &= K \Big( 1 + \frac{1}{\ln(\sqrt{B} T_K)} - \frac{1}{\ln(\sqrt{B_0} T_K)}\Big)
\end{align}
where the initial value is given by $I_{RKKY}(B_0) = K$ and $B_0 = 1/\LL_0^2$.
The above sketched calculation is only valid for $BK^2 \ll 1$ but as
we have also argued the flow for $BK^2 \gg 1$ is negligible and thus
we can write
\begin{align}
I_{RKKY}(B \to \infty) &\approx K \Big( 1 - \Big( \frac{1}{\ln(K/T_K)} - \frac{1}{\ln(\LL_0/T_K)}  \Big) \Big)
\end{align}
The calculation in this appendix illustrates that there are
terms which lead to a logarithmic correction of the exchange energy
gap. The renormalization of the quantum dot energy levels is known
and occurs for example as the Knight shift for a spin-1/2 quantum dot
in magnetic field~\cite{Fritsch:09}. 

Note that there is no further contribution to $I_{RKKY}$ from the
two-loop calculation.

\begin{widetext}

\section{two-loop}
\label{app:2loop}

The new interaction terms in $H_{int}^{(2)}$ as defined in
Eq.~\eqref{eq:Hint2} are generated due to two
Kondo spin scattering processes
\begin{align}
  \frac{d(K_{k'k;q'q}^{\isum} - K_{q'q;k'k}^{\isum})}{dB}
&= - (\e_{k'} - \e_k + \e_{q'} - \e_q)^2 (K_{k'k;q'q}^{\isum} - K_{q'q;k'k}^{\isum})
\nn   \\ &
   + \big( (\e_{k'} - \e_k) - (\e_{q'} - \e_q) \big)
     J^{\isum, j}_{k'k} J^{\isum, v}_{q'q} \nn \\ &
   + 2 \big( (\e_{k'} - \e_k + K) - (\e_{q'} - \e_q - K) \big)
     P_{k'k}^j M_{q'q}^v \nn \\ &
   + 2 \big( (\e_{k'} - \e_k - K) - (\e_{q'} - \e_q + K) \big)
     M_{k'k}^j P_{q'q}^v
\end{align}
\begin{align}
  \frac{d(K_{k'k;q'q}^{P} - K_{q'q;k'k}^{P})}{dB}
&= - (\e_{k'} - \e_k + \e_{q'} - \e_q + K)^2 (K_{k'k;q'q}^{P} - K_{q'q;k'k}^{P})
\nn    \\ &
   + \big( (\e_{k'} - \e_k) - (\e_{q'} - \e_q + K) \big)
     J^{\isum, j}_{k'k} P_{q'q}^v \nn \\ &
   + \big( (\e_{k'} - \e_k + K) - (\e_{q'} - \e_q ) \big)
     P_{k'k}^j J^{\isum, v}_{q'q}
\end{align}
\begin{align}
  \frac{d(K_{k'k;q'q}^{M} - K_{q'q;k'k}^{M})}{dB}
&= - (\e_{k'} - \e_k + \e_{q'} - \e_q - K)^2 (K_{k'k;q'q}^{M} - K_{q'q;k'k}^{M})
\nn   \\ &
   + \big( (\e_{k'} - \e_k) - (\e_{q'} - \e_q - K) \big)
     J^{\isum, j}_{k'k} M_{q'q}^v \nn \\ &
   + \big( (\e_{k'} - \e_k - K) - (\e_{q'} - \e_q ) \big)
     M_{k'k}^j J^{\isum, v}_{q'q}
\end{align}
\begin{align}
  \frac{d(K_{k'k;q'q}^{RKKY} + K_{q'q;k'k}^{RKKY})}{dB}
&= - (\e_{k'} - \e_k + \e_{q'} - \e_q)^2 (K_{k'k;q'q}^{RKKY} + K_{q'q;k'k}^{RKKY})
\nn   \\ &
   - 2 \big( (\e_{k'} - \e_k + K) - (\e_{q'} - \e_q - K) \big)
     P_{k'k}^j M_{q'q}^v \nn \\ &
   + 2 \big( (\e_{k'} - \e_k - K) - (\e_{q'} - \e_q + K) \big)
     M_{k'k}^j P_{q'q}^v
\end{align}

Exchange of summation indices provides us with a symmetry constraint
\begin{align}
  K_{k'k;q'q}^{\isum} &= - K_{q'q;k'k}^{\isum}, \\
  K_{k'k;q'q}^{P/M} &= - K_{q'q;k'k}^{P/M} ,\\
  K_{k'k;q'q}^{RKKY} &= K_{q'q;k'k}^{RKKY} .
\end{align}
In addition the hermiticity of the Hamiltonian has to be fulfilled and
such there are some simplifying relations
\begin{align}
  K_{k'k;q'q}^{\isum} &= - K_{kk';qq'}^{\isum} \\
  K_{k'k;q'q}^{P/M} &= - K_{kk';qq'}^{M/P} \\
  K_{k'k;q'q}^{RKKY} &= K_{kk';qq'}^{RKKY}
\end{align}
These relations are also fulfilled in the scaling equations.

The canonical generator $\eta^{(2)} = [H_0, H_{int}^2]$ in second
order of the Kondo coupling is explicitly given by
\begin{align}
  \eta_{int}^{(2)}
=& (\e_{k'} - \e_k + \e_{q'} - \e_q)
    K^{\isum}_{k'k;q'q} i :\big( \vec{S}_L + \vec{S}_R \big)
    \Big( \vec{s}_{(k'j)(kj)} \times \vec{s}_{(q'v)(qv)} \Big): \nn \\
&+ (\e_{k'} - \e_k + \e_{q'} - \e_q + K)
    K^{P}_{k'k;q'q} i :\Big( \big( \vec{S}_L - \vec{S}_R \big)
    + 2 i \big( \vec{S}_L \times \vec{S}_R \big) \Big)
    \Big( \vec{s}_{(k'j)(kj)} \times \vec{s}_{(q'v)(qv)} \Big): \nn \\
&+ (\e_{k'} - \e_k + \e_{q'} - \e_q - K)
   K^{M}_{k'k;q'q} i :\Big( \big( \vec{S}_L - \vec{S}_R \big)
    - 2 i \big( \vec{S}_L \times \vec{S}_R \big) \Big)
    \Big( \vec{s}_{(k'j)(kj)} \times \vec{s}_{(q'v)(qv)} \Big): \nn \\
&+ (\e_{k'} - \e_k + \e_{q'} - \e_q) K^{RKKY}_{k'k;q'q} i 2i
   :\big( \vec{S}_L \times \vec{s}_{(k'j)(kj)} \big)
    \big( \vec{S}_R \times \vec{s}_{(q'v)(qv)} \big)
   + \big( \vec{S}_R \times \vec{s}_{(k'j)(kj)} \big)
     \big( \vec{S}_L \times \vec{s}_{(q'v)(qv)} \big): \\
&=   \eta_{(2)}^\isum + \eta_{(2)}^P + \eta_{(2)}^M + \eta_{(2)}^{RKKY}
\end{align}
Note that both $K^{\isum}$ and $K^{RKKY}$ do not involve a
singlet-triplet transition like $K^{P/M}$.

From $[\eta^{(2)}, H^{(1)}]$ and $[\eta^{(1)}, H^{(2)}]$ we find the
higher order contributions to the flow $dH/dB$ and thus a scaling
equation for the coupling $g_{k'k}^{\isum,j}$ to two-loop order
\begin{align}
  \frac{d g^{\isum, j}_{k'k}}{dB} &= \ldots  \nn \\
   &+ \eh \sum_{q'q} \left[ n(q'v) (1 - n(qv)) + n(qv) (1 - n(q'v))
   \right]  
 \left( \e_{k'} - \e_k + \e_{q'} - \e_q - (\e_q - \e_{q'}) \right)
         (k_{k'k;q'q}^{\isum} - k_{q'q;k'k}^{\isum}) g^{\isum, v}_{qq'}
\nn      \\
   &+ \frac{1}{2} \sum_{q'q} (n(q'v) - n(qv))
\left( \e_{k'} - \e_k + \e_{q'} - \e_q - (\e_q - \e_{q'}) \right)
(k^{RKKY}_{k'k;q'q} + k^{RKKY}_{q'q;k'k}) g^{\isum,v}_{qq'}
\nn \\
   &+ 2 \sum_{q'q} n(qv) ( 1 - n(q'v) )
\left( \e_{k'} - \e_k + \e_{q'} - \e_q - K
            - (\e_q - \e_{q'} + K) \right)
       (k_{k'k;q'q}^{M} - k_{q'q;k'k}^{M}) p^{v}_{qq'} \nn \\
   &+ 2 \sum_{q'q} n(q'v) ( 1 - n(qv) )
\left( \e_{k'} - \e_k + \e_{q'} - \e_q + K
            - (\e_q - \e_{q'} - K) \right)
         (k_{k'k;q'q}^{P} - k_{q'q;k'k}^{P}) m^{v}_{qq'}
\end{align}
where $k_{k'k;q'q} = N(0) K_{k'k;q'q}$ and we neglected a contribution from 
\begin{align}
  - & 2\ \sum_{q'q} (n(q'v) - n(qv))
\left( \e_{k'} - \e_k + \e_{q'} - \e_q - (\e_q - \e_{q'}) \right)
  (k^{RKKY}_{k'q;q'k} + k^{RKKY}_{q'k;k'q}) g^\isum_{qq'}
  :(\vec{S}_L + \vec{S}_R) \vec{s}_{(k'j)(kj)}:
\end{align}
It can be shown in the simplest limit that this term is proportional
to $B^{-3/2}$ and thus negligible compared to the leading order
$B^{-1/2}$. In the more general cases it can be shown numerically that
the contribution from $k_{k'q;q'k}$ does not have an effect on the
flow. 

Furthermore the expression for $p_{k'k}^j$ yields
\begin{align}
  \frac{d p^{j}_{k'k}}{dB} &= \ldots \nn \\
   &+ \sum_{q'q}  n(qv) (1 - n(q'v))
     \left( \e_{q'} - \e_q + \e_{k'} - \e_k + K - (\e_q - \e_{q'}) \right)
         (k_{k'k;q'q}^{P} - k_{q'q;k'k}^{P}) g^{\isum, v}_{qq'}
\nn      \\
   &+  \sum_{q'q} n(q'v) (1 - n(qv))
      \left( \e_{q'} - \e_q + \e_{k'} - \e_k - (\e_q - \e_{q'} + K) \right)
         (k_{k'k;q'q}^{\isum} - k_{q'q;k'k}^{\isum}) p^{v}_{qq'}
\nn \\
   &- \sum_{q'q} n(q'v)(1 - n(qv))
      \left( \e_{q'} - \e_q + \e_{k'} - \e_k - (\e_q - \e_{q'} + K) \right)
    (k^{RKKY}_{k'k;q'q} + k^{RKKY}_{q'q;k'k}) p^v_{qq'}
\end{align}
where we neglect contributions from
\begin{align}
& - \sum_{q'q} [n(q'v)(1 - n(qv)) + n(qv) (1 - n(q'v))]
\Big( (\eta^{RKKY}_{k'k;q'q} + \eta^{RKKY}_{q'q;k'k}) P^v_{qq'}
  - (K^{RKKY}_{k'k;q'q} + K^{RKKY}_{q'q;k'k}) \eta^P_{qq'} \Big)
\\ & \qquad \qquad \times
:((\vec{S}_L - \vec{S}_R) + 2 i (\vec{S}_L \times \vec{S}_R)) \vec{s}_{(k'j)(kj)}:
\nn \\ &
+ 2 \sum_{q'q} n(q'v)(1 - n(qv))
\Big( (\eta^{RKKY}_{k'q;q'k} + \eta^{RKKY}_{q'k;k'q}) P^v_{qq'}
   -  (K^{RKKY}_{k'q;q'k} + K^{RKKY}_{q'k;k'q}) \eta^P_{qq'} \Big)
\\ & \qquad \qquad \times
:((\vec{S}_L - \vec{S}_R) + 2 i (\vec{S}_L \times \vec{S}_R)) \vec{s}_{(k'j)(kj)}:
\end{align}
This is a well-controlled approximation using 
the argument before that the terms proportional
to $k'q;q'k$
 can be shown numerically to be
negligible small compared to the terms of type $k'k;q'q$. Additionally
we can use the symmetry of the system that $p^L_{k'k} = - p^R_{k'k}$
and thus the first term in the latter equation cancels and can therefore be neglected.

Since we are not interested in the flow of the newly generated
couplings, we integrate out the scaling equations for them in order to
find their B-dependence. We illustrate this procedure on $K^\isum_{k'kq'q}$ 
as an example:
\begin{align}
  (K_{k'k;q'q}^{\isum} - K_{q'q;k'k}^{\isum})(B)
&= \ee^{- B (\e_{k'} - \e_k + \e_{q'} - \e_q)^2} 
\int\limits_{B_0}^B dB' \ee^{+ B' (\e_{k'} - \e_k + \e_{q'} - \e_q)^2} 
\nn \\ & \qquad 
\Big\{ 
   \big( (\e_{k'} - \e_k) - (\e_{q'} - \e_q) \big)
     J^{\isum, j}_{k'k}(B')  J^{\isum, v}_{q'q}(B')  \nn \\ & \qquad
   + 2 \big( (\e_{k'} - \e_k + K) - (\e_{q'} - \e_q - K) \big)
     P_{k'k}^j (B')  M_{q'q}^v (B')  \nn \\ & \qquad
   + 2 \big( (\e_{k'} - \e_k - K) - (\e_{q'} - \e_q + K) \big)
     M_{k'k}^j (B')  P_{q'q}^v (B')   
\Big\}
\end{align}
In the following we apply the
diagonal parametrization for the Kondo couplings and thus find
\begin{align}
  (K_{k'k;q'q}^{\isum} - K_{q'q;k'k}^{\isum})(B)
&= \ee^{- B (\e_{k'} - \e_k + \e_{q'} - \e_q)^2} 
\int\limits_{B_0}^B dB' \ee^{+ B' (\e_{k'} - \e_k + \e_{q'} - \e_q)^2} 
\nn \\ & \qquad 
\Big\{ 
   \big( (\e_{k'} - \e_k) - (\e_{q'} - \e_q) \big)
    \ee^{- B' (\e_{k'} - \e_k)^2} J^{\isum, j}_{k}(B') 
    \ee^{- B' (\e_{q'} - \e_q)^2} J^{\isum, v}_{q}(B')  \nn \\ & \qquad
   + 2 \big( (\e_{k'} - \e_k + K) - (\e_{q'} - \e_q - K) \big)
     \ee^{- B' (\e_{k'}- \e_k + K)^2} P_{k}^j (B')  
     \ee^{- B' (\e_{q'} - \e_q - K)^2} P_{q}^v (B')  \nn \\ & \qquad
   + 2 \big( (\e_{k'} - \e_k - K) - (\e_{q'} - \e_q + K) \big)
     \ee^{- B' (\e_{k'}- \e_k - K)^2} P_{k}^j (B')  
     \ee^{- B' (\e_{q'} - \e_q + K)^2} P_{q}^v (B')  
\Big\}
\end{align}
In the calculation to one-loop order we found that the Kondo couplings
depend logarithmically on the flow parameter and thus have a much
slower dependence than the exponential function in the latter
expression. In the following we therefore assume that we can replace
the coupling by their averaged value in the interval from $B_0$ to
$B$. In general the initial value $B_0= 1/\LL_0^2$ should be chosen as
the band cutoff in order to treat boundary terms. Since this is
equivalent to treating a different model with new boundary conditions
we set in the following as also in the main text
\begin{align}
  \label{eq:9}
  B_0 = 0
\end{align}
Using these two simplifications the integration is simple and can be
done straightforwardly yielding
\begin{align}
  (K_{k'k;q'q}^{\isum} - K_{q'q;k'k}^{\isum})(B)
&= \frac{1}{2B} \ee^{- B (\e_{k'} - \e_k + \e_{q'} - \e_q)^2} \nn \\ & \qquad
\Big\{ 
\frac{  (\e_{k'} - \e_k) - (\e_{q'} - \e_q) }{(\e_{k'} - \e_k) (\e_{q'} - \e_q)}
\big( \ee^{2 B (\e_{k'} - \e_k)(\e_{q'} - \e_q)} - 1 \big)
\Big[ \frac{1}{B} \int\limits_{0}^B dB'  J^{\isum, j}_{k}(B')
J^{\isum, v}_{q}(B')  \Big] \nn \\ & \qquad
   + 2 
    \frac{ (\e_{k'} - \e_k + K) - (\e_{q'} - \e_q - K) }{ (\e_{k'} - \e_k + K)  (\e_{q'} - \e_q - K)}
\big( \ee^{2 B (\e_{k'} - \e_k + K)(\e_{q'} - \e_q - K)} - 1 \big)
\Big[ \frac{1}{B} \int\limits_{0}^B dB'  P_{k}^j (B')   P_{q}^v (B')
\Big] \nn \\ & \qquad
   + 2 
    \frac{ (\e_{k'} - \e_k - K) - (\e_{q'} - \e_q + K) }{ (\e_{k'} - \e_k - K)  (\e_{q'} - \e_q + K)}
\big( \ee^{2 B (\e_{k'} - \e_k - K)(\e_{q'} - \e_q + K)} - 1 \big)
\Big[ \frac{1}{B} \int\limits_{0}^B dB'  P_{k}^j (B')   P_{q}^v (B')  \Big]
\Big\}
\end{align}

The same procedure is applied to the other coupling generated to
two-loop order and these expressions are then inserted in the flow
equation for the Kondo coupling. In the following we write down only
the result for the diagonal parametrization where we assume e.g.~in
$g_k^{\isum,j}$ that $\e_{k'} = \e_k$. We find that the collected terms
contributing to $dg_k^{\isum,j}/dB$ in
two-loop order can be written by $TL(q,q')+TL(q',q)$ where
\begin{align}
TL(q,q') &=
     \eh B \sum_{q'q} n(q'v) (1 - n(qv)) 
\big( - 2 (\e_{q'} - \e_q)^2 \big)
     \ee^{- 2 B (\e_{q'} - \e_q)^2}
\nn   \\ & \qquad
     \Big[ \frac{1}{B} \int_{B_0}^B dB' g_{k}^{\isum, j}(B')
     g_{(q'+q)/2}^{\isum, v}(B') \Big]
     g^{\isum, v}_{(q+q')/2}(B)
     \nn \\
   &+ 2 B \sum_{q'q} n(q'v) ( 1 - n(qv) )
         \big(- 2 (\e_{q'} - \e_q + K)^2 \big)
         \ee^{- 2 B (\e_{q'} - \e_q + K)^2}
\nn \\ & \qquad
     \Big[ \frac{1}{B} \int_{B_0}^B dB' g_{k}^{\isum, j}(B')
     p_{(q'+q)/2}^v(B') \Big]
     p^{v}_{(q+q')/2}(B)
      \nn   \\
&+
     2 \sum_{q'q} \big[ n(q'v) (1 - n(qv)) \big]
         \ee^{- 2 B (\e_{q'} - \e_q)^2}
     \frac{(\e_{q'} - \e_q + K)^2 - K^2}{K(\e_{q'} - \e_q + K)}
     \big( \ee^{2 B (- K)(\e_{q'} - \e_q + K)} - 1\big)
\nn     \\ & \qquad
     \Big[ \frac{1}{B} \int_{B_0}^B dB' p_{k}^{j}(B')
     p_{(q'+q)/2}^v(B') \Big]
         g^{\isum, v}_{(q+q')/2}(B)
\nn     \\
   &+   2 \sum_{q'q} \big[ n(q'v) (1 - n(qv)) \big]
       \ee^{- 2 B (\e_{q'} - \e_q + K)^2}
      \frac{K^2 - (\e_{q'} - \e_q)^2}{K(\e_{q'} - \e_q)}
    (\ee^{2 B (+ K)(\e_{q'} - \e_q)} - 1)
\nn \\ & \qquad
     \Big[ \frac{1}{B} \int_{B_0}^B dB' p_{k}^{j}(B')
     g^{\isum, v}_{(q'+q)/2}(B') \Big]
     p^{v}_{(q+q')/2}(B)
\end{align}
Note that $TL(q,q') = TL(q',q)$.
For the symmetric setup we are studying, only the first two terms are
non-zero after summation over the lead index ($p_k^L = - p_k^R$)
and the last two terms are thus neglected in the following. 
Otherwise the last two terms are nasty whereas the first two terms can
be summed over momentum straightforwardly.

Please note that the last two terms are of the same order than the
first two terms if $K = 0$. In the expressions which are 
discussed in the main text it is thus not valid to set $K = 0$,
especially in the case when one wants to derive the spin-1/2 limiting
case where the cancellation of $p_k^L = - p_k^R$ does not take place.

In $dp/dB$ the terms are less symmetric
\begin{align}
\frac{d\, p_{k}^{j}(B)}{dB} &= \text{one-loop} 
\nn \\
   &+ B \sum_{q'q}  n(qv) (1 - n(q'v))
     \left( - 2 (\e_{q'} - \e_q)^2  \right)
     \ee^{- 2 B(\e_{q'} - \e_q)^2}
\nn     \\ & \qquad
     \Big[ \frac{1}{B} \int_{B_0}^B dB'
     p_{k}^{j}(B')
     g^{\isum, v}_{(q'+ q)/2}(B') \Big]
     g^{\isum, v}_{(q+q')/2}
\nn     \\
   &+ 4 B \sum_{q'q} n(q'v)(1 - n(qv))
      \big( - 2 (\e_{q'} - \e_q - K)^2 \big)
      \ee^{- 2 B(\e_{q'} - \e_q - K)^2}
\nn \\ & \qquad
  \Big[ \frac{1}{B} \int_{B_0}^B dB'
           p_{k}^{j}(B')
           p_{(q'+q)/2}^{v}(B') \Big]
           p^{v}_{(q+q')/2}
\nn \\
   &+ \sum_{q'q} n(q'v) (1 - n(qv))
\ee^{- 2 B (\e_{q'} - \e_q - K)^2}
   \frac{(\e_{q'} - \e_q)^2 - K^2}{K (\e_{q'} - \e_q)}
   \big( \ee^{2 B(- K)(\e_{q'} - \e_q)} - 1 \big)
\nn \\ & \qquad
   \Big[ \frac{1}{B} \int_{B_0}^B dB'
     g_{k}^{\isum, j}(B')
     g_{(q'+q)/2}^{\isum, v}(B') \Big]
     p^{v}_{(q+q')/2}
\nn \\
   &+ \sum_{q'q}  n(qv) (1 - n(q'v))
     \ee^{- 2 B(\e_{q'} - \e_q)^2}
     \frac{(\e_{q'} - \e_q + K)^2 - K^2}{K (\e_{q'} - \e_q + K)}
     \big( \ee^{2 B (-K)(\e_{q'} - \e_q + K)} - 1 \Big)
\nn     \\ & \qquad
     \Big[ \frac{1}{B} \int_{B_0}^B dB'
     g_{k}^{\isum, j}(B')
     p_{(q'+q)/2}^v(B') \Big]
     g^{\isum, v}_{(q+q')/2}
\end{align}
\end{widetext}
Note that we do not have $q \leftrightarrow q'$ symmetry! This is
somehow expected since $p_{k'k}$ is only symmetric under exchange of
$k \leftrightarrow k'$ and also $K \leftrightarrow - K$. The latter
symmetry is fulfilled in the scaling equation.

For the energy integration the following relation is useful
\begin{align}
   & \int d\e_{q'} \int d\e_q n(qv)(1 - n(q'v)) 
\\ & \qquad \qquad \times
       (\e_{q'} - \e_q + \alpha K)^2 
       \mathrm{e}^{- 2 B \left( \e_{q'} - \e_q + \alpha K \right)^2}
       \nn \\
  &= 
  \big( 1 - \frac{2 r_v}{(1 + r_v)^2} \big) F(\alpha K) 
\\ & \qquad 
  + \frac{r_v}{(1 + r_v)^2} \left( F(\alpha K + eV_v) 
                                  + F(\alpha K - eV_v) \right)
\end{align}
where contributions from the band cutoff cancel out or are exponentially small
and 
\begin{align}
  F(y) &= \frac{y}{8 B} \frac{\sqrt{\pi}}{\sqrt{2 B}}  \,
           \mathrm{erf}\left(\sqrt{2B} \, y \right)
        + \frac{1}{2 (2 B)^2} \, \mathrm{e}^{- 2 B y^2}.
\end{align}
The explicit expression for $dg^{\isum,j}_{k}/dB$ after integration is given in the main text.

\begin{widetext}
\begin{align}
\frac{d\, g_{k}^{\isum, j}(B)}{dB}
 &= \quad
\frac{1}{2B} \Big( \frac{r_j}{1+r_j} \ee^{-2 B (\e_{k} - V_j/2)^2} +
\frac{1}{1+r_j} \ee^{-2 B (\e_{k} + V_j/2)^2} \Big)
  \big( g_{k}^{\isum, j} \big)^2
\nn \\ & \quad
+ 2 \frac{1}{2B}
  \Big( \frac{r_j}{1+r_j}\ee^{-2 B(\e_{k} -V_j/2 + K)^2}
  + \frac{r_j}{1+r_j}\ee^{-2 B(\e_{k} -V_j/2 + K)^2}  \Big)
  \big( p_{k+K/2}^{j} \big)^2
\nn \\ & \quad
+ 2  \frac{1}{2B}
  \Big( \frac{r_j}{1+r_j} \ee^{-2 B(\e_{k} - V_j/2 - K)^2}
      + \frac{1}{1+r_j} \ee^{-2 B(\e_{k} + V_j/2 - K)^2} \Big)
  \big( p_{k-K/2}^{j} \big)^2
\nn \\ & \quad
   - \sum_v\  \frac{1}{4B}  
    g_{k}^{\isum, j}(B)
     \big( g_{max}^{\isum, v}(B) \big)^2
\nn \\ & \quad
   -   \sum_v\  g(0, r_v, V_v)
    g_{k}^{\isum, j}(B)
     \big( g_{max}^{\isum, v}(B) \big)^2
\nn \\ &\quad
   -   \sum_v\ \big( \frac{1}{4 B} \ee^{- 2 B K^2} + \frac{K}{4} \sqrt{\frac{\pi}{2B}} \mathrm{erf}\big(
  \sqrt{2B} K\big) \big)
    g_{k}^{\isum, j}(B)
   \big( 2   p_{max}^v(B) \big)^2
\nn \\ &\quad
   -   \sum_v\  g(K, r_v, V_v)\ 
    g_{k}^{\isum, j}(B)
   \big(  2 p_{max}^v(B) \big)^2
\end{align}
and
\begin{align}
\frac{d\, p_{k}^{j}(B)}{dB} &= \frac{1}{2B}
  \big( \frac{r_j}{1+r_j} \ee^{- 2 B (\e_{k} - V_j/2 - K/2)^2}
  + \frac{1}{1+r_j} \ee^{- 2 B (\e_{k} + V_j/2 - K/2)^2} \big)
  g_{k-K/2}^{\isum, j}
  p_{k}^{j}
\nn \\ & \quad
+ \frac{1}{2B}
  \big( \frac{r_j}{1+r_j} \ee^{- 2 B (\e_{k} - V_j/2 + K/2)^2}
  + \frac{1}{1+r_j} \ee^{- 2 B (\e_{k} + V_j/2 + K/2)^2} \big)
  p_{k}^{j}
  g_{k+K/2}^{\isum, j}
\nn \\   & \quad
    - \sum_v \ \frac{1}{4B}  
    p_{k}^{j}(B)
    \big(  g^{\isum, v}_{max}(B) \big)^2
\nn \\   & \quad
    -  \sum_v\  g(0, r_v, V_v) 
    p_{k}^{j}(B)
    \big(  g^{\isum, v}_{max}(B) \big)^2
\nn \\   & \quad
     -  \sum_v\ \big( \frac{1}{4 B} \ee^{- 2 B K^2} + \frac{K}{4} \sqrt{\frac{\pi}{2B}} \mathrm{erf}\big(
  \sqrt{2B} K\big) \big)
          p_{k}^{j}(B)
          \big( 2 p_{max}^{v}(B) \big)^2
\nn \\   & \quad
     -  \sum_v\  g(K, r_v, V_v)
          p_{k}^{j}(B)
          \big( 2 p_{max}^{v}(B) \big)^2
\end{align}
where
\begin{align}
g(K,r,V) &= \frac{1}{4 B} \frac{r}{(1+r)^2} \big( \ee^{- 2 B (K-V)^2} + \ee^{-
  2 B (K + V)^2} - 2 \ee^{- 2 B K^2} \big) \nn \\
&+ \frac{1}{4} \sqrt{\frac{\pi}{2B}} \frac{r}{(1+r)^2} 
  \big( (K-V) \mathrm{erf}(\sqrt{2B} (K-V))
+ (K+V) \mathrm{erf}(\sqrt{2B} (K+V))
- 2 K \mathrm{erf}(\sqrt{2B} K) \big)  
\end{align}
For $r = 1$ the function $g(K,V)$ corresponds to the function $f(K,V)$
in the main text.

\end{widetext}

\end{document}